\documentclass[
preprint,
bibnotes,
amsmath,amssymb,
]{revtex4-2}

\usepackage{graphicx}
\usepackage{dcolumn}
\usepackage{bm}
\usepackage[normalem]{ulem}
\usepackage{makecell}
\usepackage{amsmath}
\usepackage{colonequals}
\usepackage{mathtools}

\DeclarePairedDelimiter\floor{\lfloor}{\rfloor}

\begin{document}

\preprint{APS/123}

\title{An Enquiry on similarities between Renormalization Group and Auto-Encoders using Transfer Learning.}

\author{Mohak Shukla}
  \email{1821ph07@iitp.ac.in}
\author{Ajay D. Thakur}%
 \email{ajay.thakur@iitp.ac.in}
\affiliation{%
 Department of Physics, \\Indian Institute of Technology Patna, \\Bihta 801106, India
}%

\date{\today}

\begin{abstract}
Physicists have had a keen interest in the areas of Artificial Intelligence (AI) and Machine Learning (ML) for some time now, with a special inclination towards unravelling the mechanism at the core of the process of learning. In particular, exploring the underlying mathematical structure of a neural net (NN) is expected to not only help us in understanding the epistemological meaning of ‘Learning’ but also has the potential to unravel the secrets behind the workings of the brain. Here, it is worthwhile to establish correspondences and draw parallels between methods developed in core areas of Physics and the techniques developed at the forefront of AI and ML. Although recent explorations indicating a mapping between the Renormalisation Group(RG) and Deep Learning(DL) have shown valuable insights, we intend to investigate the relationship between RG and Autoencoders(AE) in particular. We will use Transfer Learning(TL) to embed the procedure of coarse-graining in a NN and compare it with the underlying mechanism of encoding-decoding through a series of tests.
\begin{description}
\item[Keywords]
Renormalization Group; Deep Learning; Auto-Encoders; Transfer Learning; Ising model; Unsupervised Learning.
\end{description}
\end{abstract}

\maketitle

\section{\label{sec:Intro}Introduction}

Machine Learning or as some of us would call it, Statistical-Optimization is not a new field per se. Conceptualization of Artificial Intelligence(AI) (of which Machine Learning(ML) is a part) began as soon as computers became capable of executing a complex set of instructions. Hence the words Artificial Intelligence and Machine Learning were coined in the 1950s and 1960s, respectively \cite{Mccarthy1955, Samuel1959}.The field, however, remained mostly in the conceptual realm \cite{little1974, hopfield1982, amit1985, amit1992} due to a lack of computational power, which was needed to scale the algorithms to solve any worthwhile problem. In the last couple of decades, however, the advancement in silicon fabrication and multi-core processors has reinvigorated the field, making it possible to crunch large sets of data on training machine learning models. Today ML is used to solve various kinds of problem, from object detection to high frequency stock-trading.\par
\begin{figure}[h]
{\includegraphics[width=.9\linewidth]{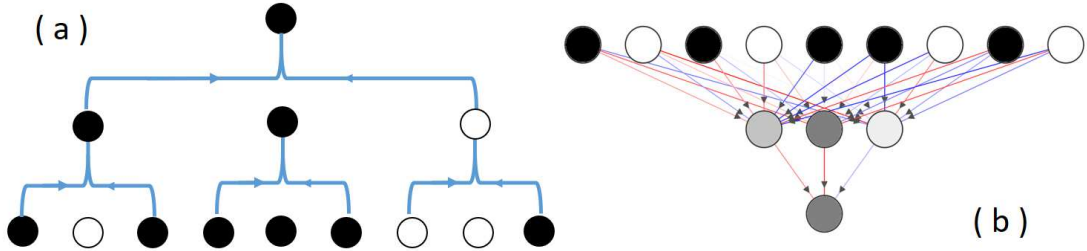}}
  \caption{\label{fig:RG NN} Showing an example of (a) Coarse-graining used in RG for 1-dimensional Ising model using 3-spinblock majority rule. (b) A basic NN consisting of input, hidden and output layers.}
\end{figure}

By the end of the 20th century, physicists have found themselves surrounded by problems that are inconceivably hard to solve. Equations governing millions of stars in a galaxy are simply not solvable by any known analytical method; the same goes true for solving the wave-function of exotic particles in a high-energy physics problem and computational physics appears to be the only way forward. This is where ML can be of immense assistance \cite{Teimoorinia2012, decelle2019, umrigar1988, ch2017, torlai2016}. On the other hand, physicists have also been contributing to ML, drawing inspiration from the theories and techniques developed for computational physics, as well as providing insights into the foundation of AI and ML \cite{shah2019, Tegmark2017, amin2018, puvskarov2018}. One such insight manifests as we study deep learning side by side with the renormalization group (Fig.\ref{fig:RG NN}).\par

An artificial neuron was designed to mimic the workings of a biological neuron using mathematical functions \cite{mcculloch1943}. Rosenblatt used them to make a network that can learn to act as a binary classifier, called it the Perceptron \cite{rosenblatt1958}. After a while these neural-nets were stacked on top of each other to make a multi-class discriminator, such NNs were called multilayer perceptron(MLP) or deep neural network(DNN). DNN are known for their ability to extract features(relations) from the dataset it is trained on. These features become more complex as we go deeper inside the layers of a neural net. For example, a DNN trained on Human faces may recognize the outlines of a face in the first layer. Going a layer deeper, it might start recognizing facial features like eyes, nose, lips. A layer deeper, it might start understanding the relative positioning of eyes/nose/lips in order to recognize and differentiate faces. This reminds us of the coarse-graining techniques employed in RG \cite{Stueckelberg1953, Wilson1983}. Coarse-graining is essentially summing/integrating over short-range interactions. Resulting in long-range relations and fewer degrees of freedom. RG thus gives us a systematic procedure to determine the macroscopic relations governing the system from microscopic interactions, iteratively. Indicating an analogy with DL. Some pioneering work has been done in this area \cite{Li2018, beny2013} particularly the work on finding an exact mapping between the Variational Renormalization Group and Deep Learning by Mehta and Schwab \cite{mehta2014}. We, however, see room for further exploration which will refine our understanding of the subject.\par
Thus we propose a reductionist approach to investigate the parallels between RG and DL. For that, we intend to dissect both the processes down to the individual steps, fine-tuning them to fit a common NN architecture. This will enable us to study the structures formed in the net as it learns to compress the Ising lattices and compare it with mapping formed by coarse-graining through RG. For this, we are going to prepare 3 models of NNs, each having the same number of layers, neurons, and activation function(s). The difference lies in their training routine, where one will use Backpropagation to learn encoding-decoding(i.e. an Autoencoder) while the other has a hard-coded map of decimation to perform coarse-graining and so on(see Sec.\ref{sec:enquiry} and table \ref{table:NNs}). We will also benchmark their performance on metrics such as information loss, change in entropy, the flow of inverse temperature($\beta$) and fixed point(s) which will further help in demonstrating parallels (and differences) between RG and DL in Sec. \ref{sec:Results}.

\section{\label{sec:enquiry}Method of Enquiry}

We will be using Python version 3.6 along with libraries such as Keras \cite{keras} which runs over the framework of Tensorflow \cite{tensorflow} to build and train NNs. We utilise NumPy \cite{NumPy} for matrix operation, Scikit-image \cite{Ent_skimage} image processing and Matplotlib \cite{Matplotlib} for plotting graphs. The project environment was configured to use the CUDA \cite{cuda} framework in order to use Nvidia GPUs for faster training.\par

\subsection{\label{sec:RG_deci}Renormalization through Decimation}

\begin{figure}[h]
{\includegraphics[width=.45\linewidth]{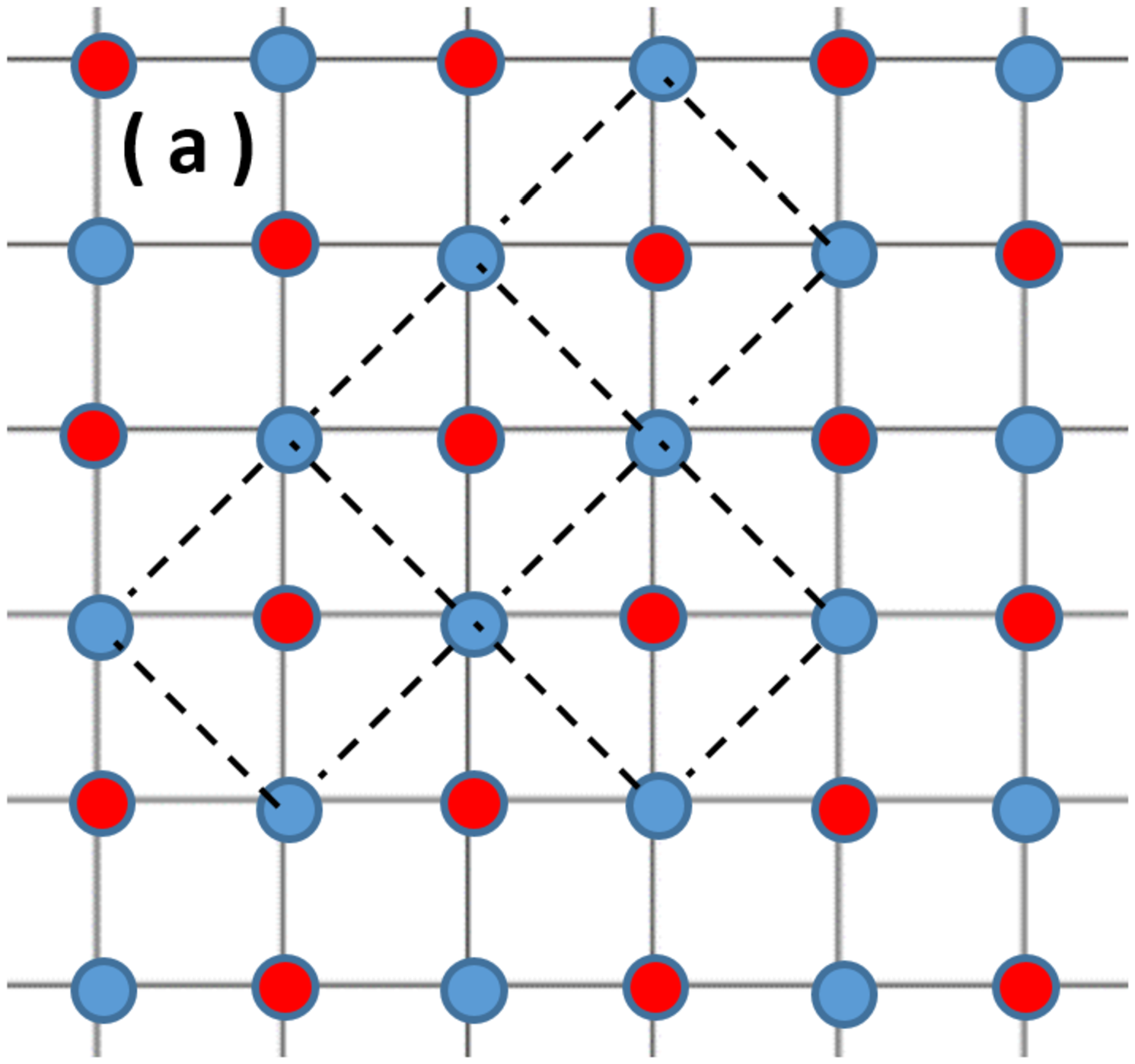}}%
{\includegraphics[width=.4\linewidth]{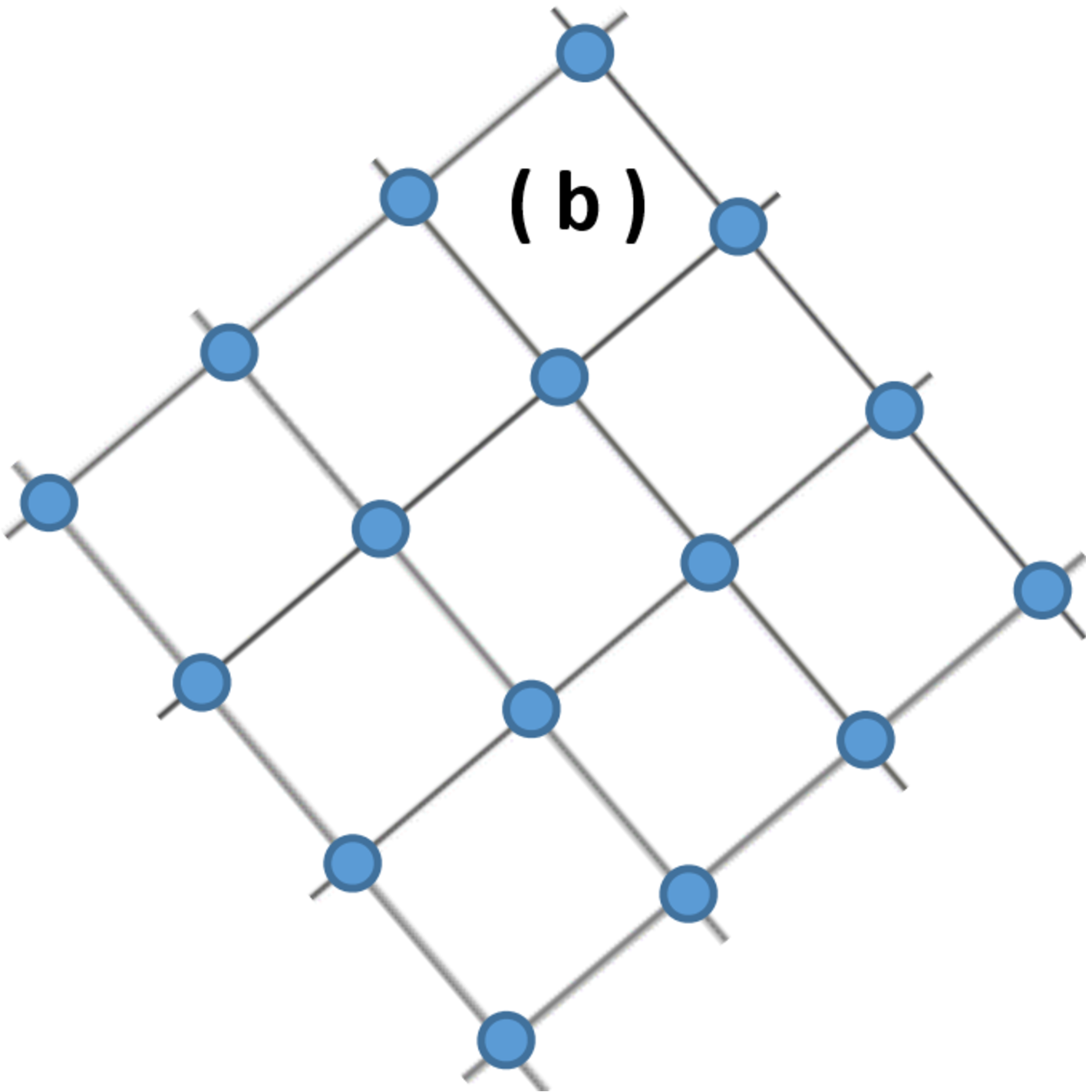}}
  \caption{\label{fig:Lat_deci} The decimation scheme:
(a) The original lattice with alternative Blue and Red spins. Red spins are then removed.
(b) The coarse-grained lattice with the emergent nearest neighbors, shown as dashed lines in (a)}
\end{figure}
In a 2-dimensional Ising lattice, each site `i' has a spin denoted by variable $(\sigma_i)$, which has values ($+1$) or ($-1$) \cite{Ising1925}.
The Hamiltonian `$H(\sigma)$' of such system is defined as:
\begin{equation}
   H(\sigma) = -J\sum_{<i,j>} \sigma_i\sigma_j
\end{equation}
For RG, we would be employing coarse-graining through Decimation. As illustrated in Fig.\ref{fig:Lat_deci}, we try to remove every other lattice site by summing over the effects of decimated Red-spins(R) to form a new lattice with only Blue-spins(B). The configuration probability in terms of inverse temperature($\beta$) is:
\begin{equation}
  P(\sigma) = \frac{e^{-\beta H(\sigma)}}{Z(\beta)} \text{, where partition function } Z(\beta) = \sum_{\sigma} e^{-\beta H(\sigma)}
\end{equation}

Expanding the term for a single decimation site (A):
\begin{equation}
  P(\sigma) = \frac{e^{\beta(B_1R_A+B_2R_A+B_3R_A+B_4R_A)}\times B}{Z(\beta)}
\end{equation}
Where  $B = exp(\beta \sum_{i,j\neq A} J_{i,j}B_iR_j) $  and the minus sign is absorbed into '$J$'.
Averaging out the effect of ($R_A$) for both (+1 and -1 spin) turns out to be:
\begin{equation}
  P(\sigma) = B \times \frac{ e^{\beta(B_1+B_2+B_3+B_4)}+e^{-\beta(B_1+B_2+B_3+B_4)}}{Z(\beta)}
\end{equation}
Expanding the term averages out the effect of spin at site(A) into pair-wise and quartet interactions of nearest and next-nearest spins in the new lattice. Taking approximation with only pair-wise coupling, without changing the Hamiltonian, while negating the quartet couplings. The new configuration probability can be approximated as:

\begin{equation}
\label{eq:P_final}
  P'(\sigma) = \frac{e^{\beta' H(\sigma)}}{Z(\beta')} \quad \text{, where $\beta' = \frac{3}{8}\ln(\cosh(4\beta))$}
\end{equation}

\subsection{\label{sec:AE}Autoencoder and the common Architecture}

\begin{figure}[ht]
{\includegraphics[width=\linewidth]{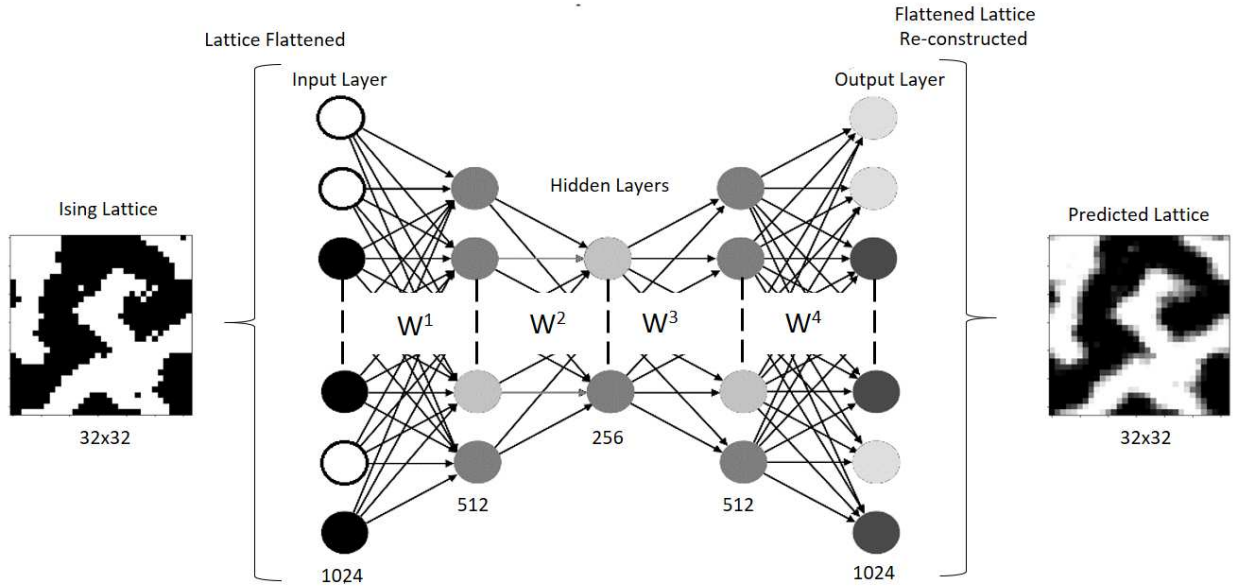}}
  \caption{\label{fig:MLP} Showing the  workings of Autoencoder used to compress the Ising lattice. This architecture will remain common for the NNs used in all three models, each consisting one input layer, three hidden layers and one output layer (further explained in Sec.\ref{sec:training} and Table \ref{table:NNs}) .}
\end{figure}

Autoencoder is a simple NN that learns to reconstruct its input by first compressing(encoding) it down to a reduced representation and then decompressing(decoding) it to get something as close to the original input as possible \cite{Rumelhart1986}\cite{Hinton2006}. This encoding is achieved by finding and reinforcing the hierarchy of relations present in the input dataset until only the ‘relevant’ features remain(Fig. \ref{fig:MLP}). This is done by tuning the weights which connect the layers to reduce losses through back-propagation \cite{LeCun1989}. Since the training dataset need not be labelled, the process is called unsupervised learning(UL).
The number of neurons in the input layer($N_I$) will equal to the number of spins in the square spin-lattice of size ($L \times L$), for $L=32$, $L^2 = 1024 = N_I$. In order to remain homologous to the decimation coarse graining, next two encoding hidden layer will be of size $N_{H1}=512$ and $N_{H2}=256$ respectively. Similarly, decoding layer $N_{H3}=512$ and finally the output layers $N_{O}=1024$. \emph{This configuration of layers will remain common for all three models.}

\subsection{\label{sec:TL}Transfer Learning}
\begin{figure}[h]
{\includegraphics[width=.9\linewidth]{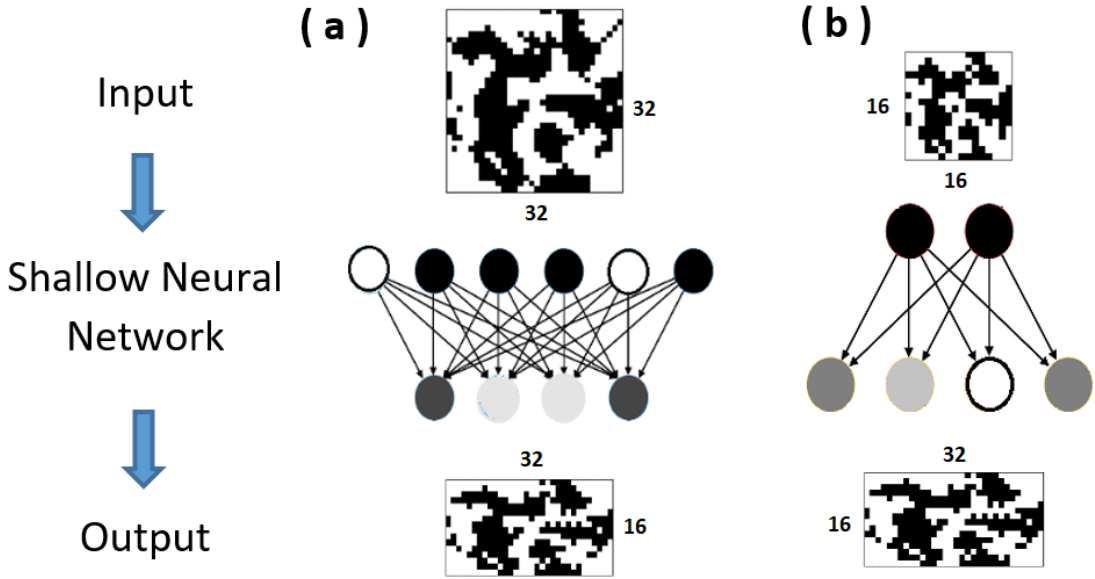}}
  \caption{\label{fig:encoder-decoder} Shallow Networks performing (a)Encoding the $32\times32$ Ising lattice down to $32\times16$  and (b)Decoding the $32\times16$ Ising lattice up to $32\times32$.\\  }
\end{figure}
Transfer-learning(TL) is a technique which uses learning from one dataset and applying it to related but different problem after some modifications. It is essentially a transfer of weights of trained layers from one NN to another, thus skipping most of training required in the second NN. For example, a NN trained to recognize cats, has also learned to recognize 4-legged animals in its hidden layers, this learning can then be transferred with slight modification to a new NN, which can recognize dogs.
In our case, several shallow-nets with just the input and the output layer will be trained to perform a single iteration of RG.\par
In the Fig.\ref{fig:encoder-decoder} two such shallow-nets are shown where former is trained to perform coarse graining (Let us call them CG) and the latter is trained to perform upscaling (UP). The input as well the output of these shallow-nets are linearized Ising-lattices. It is important to note that a single iteration of decimation reduces the number of sites in the lattice by half, and since we cannot reduce both dimensions of the lattice by the factor of $\sqrt{2}$, the reconstructed lattices will appear compressed in 1 dimension while the other remains unchanged. For example in Fig. \ref{fig:encoder-decoder}(a), a $32\times32$ Lattice is linearized to $1024$ sized array, after one decimation it will be reduced to an array of size $512$, this can only be reconstructed into a lattice of $32\times16$ or $16\times32$, hence the deformed lattice. No such problem exists in even number of iterations, where both dimensions of the lattice get reduced by the factor of 2, see the input lattice of $16\times16$ in Fig.\ref{fig:encoder-decoder}(b) for example. We are going to need four such shallow-nets:
\begin{enumerate}
  \item	Coarse-grainer from 32x32 to 32x16 (CG-1024-512)
  \item Coarse-grainer from 32x32 to 32x16 (CG-512-256)
  \item Up-scaler from 16x16 to 32x16 (UP-256-512)
  \item Up-scaler from 32x16 to 32x32 (UP-512-1024)
\end{enumerate}

\begin{figure}
\includegraphics[width=.9\linewidth]{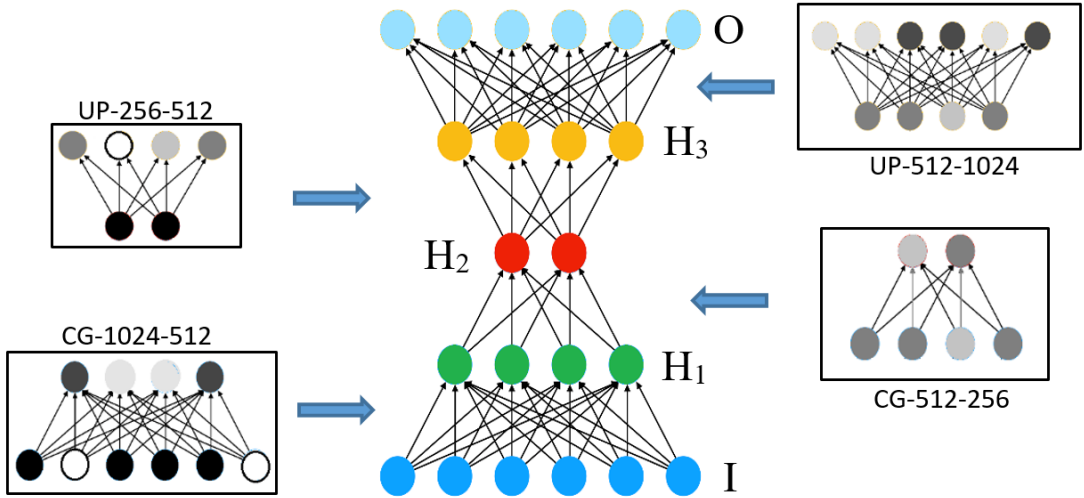}
\caption{\label{fig:TL} Showing the process of inserting shallow-nets into the frame of an untrained NN through Transfer Learning.}
\end{figure}

After training, their weights will be transferred to a multilayered NN (Fig.\ref{fig:TL}). This will enable us to construct a curated DNN, which would be similar to the Autoencoder in structure but had acquired some (or all) of its training through transfer-learning. Since the activation-function is linear and biases are not present, the output(Y) by $n^{th}$ layer is given simply by

\begin{equation}
  Y_{j}^{(n)} = \sum_{i=1}^{N} Y_{i,j}^{(n-1)}W_{i,j}^{(n-1)}
\end{equation}
where $W^{(n-1)}$ is the weight matrix connecting $(n-1)^{th}$ layer with $(n)^{th}$ layer and `N' is the number of neurons/nodes in $(n-1)^{th}$ layer.

\subsection{\label{sec:hard-code}Hard-coding the coarse graining map}
\begin{figure}[h]
{\includegraphics[width=.9\linewidth]{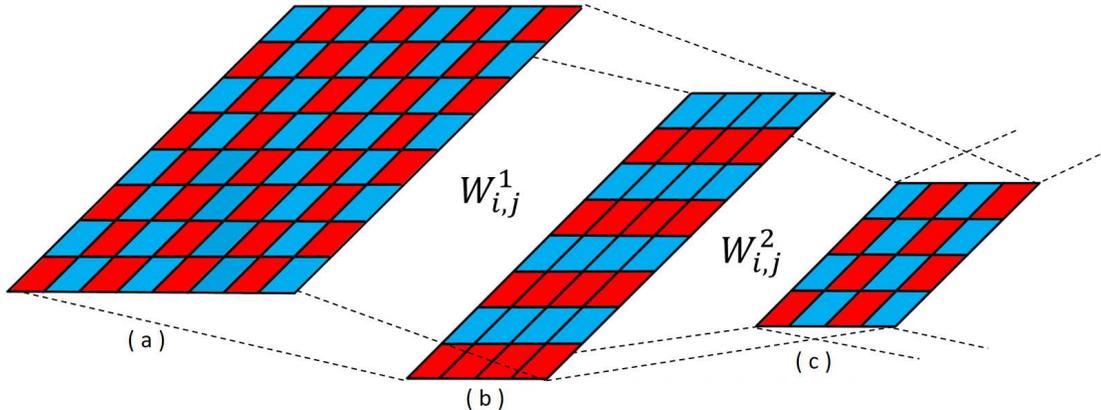}}
  \caption{\label{fig:deci_scheme} Scheme of hard-coding the decimation. Every other site (in Red) is removed and only Blue sites are carried forward into the next layer.}
\end{figure}
Techniques used to course-grain the system are relations which map the lattice from a higher dimension to lower. This map can be translated in into a graph, which can then act as a NN where weights are hard-coded instead of trained. This Hard-coded NN will act as a benchmark across which we will compare the performance of other models.
As discussed above the lattice has to be linearized and decimation has to be mapped on the weight matrices($W_{i,j}$) keeping Numpy library's linearizing scheme in mind. Equations \ref{eq:w1} and \ref{eq:w2} show the sparse matrices $W_{i,j}$ as function of indices (i,j), following the scheme which will mimic RG through decimation in a linearized lattice.
\begin{equation}
  \label{eq:w1}
    W_{i,j}^1 \colonequals
    \begin{cases}
      1 & \text{if }  \floor{\frac{i}{32}+i} \text{ is even} \\
      0  & \text{else}
    \end{cases}
\end{equation}
\begin{equation}
  \label{eq:w2}
    W_{i,j}^2 \colonequals
    \begin{cases}
      1 & \text{if } \floor{\frac{i}{16}} \text{ is even} \\
      0  & \text{else}
    \end{cases}
\end{equation}
Shallow-nets with hard-coded coarse-graining will be very similar to the trained shallow-nets mentioned above, but will carry a prefix `H' in their name, for example: Coarse-grainer from $32\times32$ to $32\times16$ will be called \textbf {H}CG-1024-512.

\subsection{\label{sec:dataset}Preparing the Dataset}
\begin{figure}[h]
{\includegraphics[width=.9\linewidth]{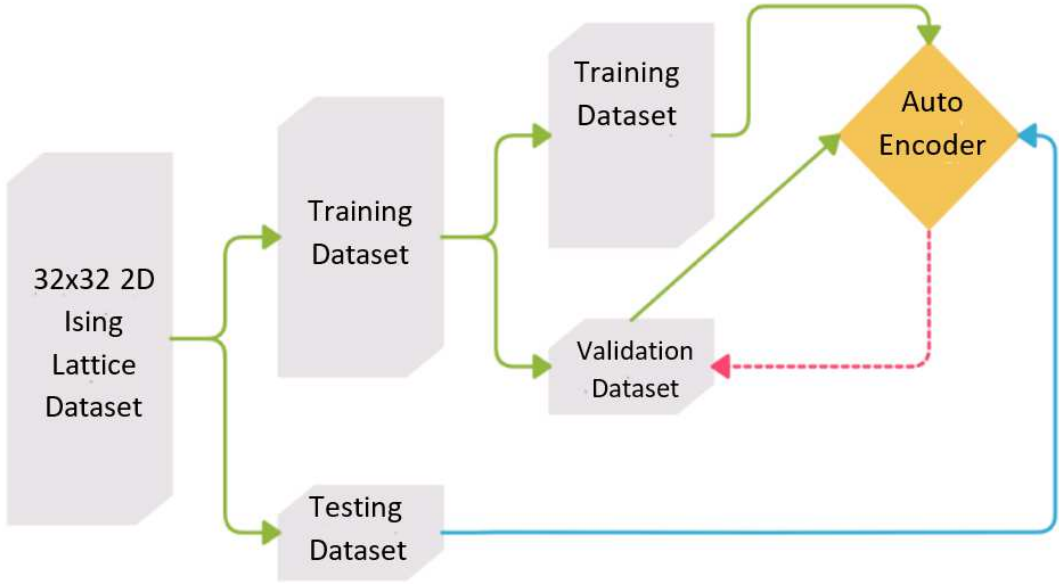}}
  \caption{\label{fig:dataset} Scheme of classifying the training data into different labelled-sets.}
\end{figure}
Data samples are generated from a Monte-Carlo simulation of 2D Ising model for $ \beta = 0.4352$, near the critical temperature. 30,000 samples of $32\times32$ sized Ising model lattice with periodic boundary condition along 2 lattices of consecutive decimations of each sample, bringing the total to 90,000 samples (see Fig.\ref{fig:Lat_deci}). Additionally 50,000 spin configurations of size $32\times32$ were generated for various $ \beta \in [0.01,0.6)$ to train inverse-temperature predicting NN and a separate auto-encoder based on Model-1 (explained in section \ref{sec:entropy} and \ref{sec:RG_flow}). These sets are then bifurcated into Training and Testing Datasets respectively. Training datasets are further bifurcated into Training and Cross-validation datasets. A simple flowchart of dataset classification is shown in the Fig. \ref{fig:dataset}.

\subsection{\label{sec:training}Training}

Some common features among all the NN (both Shallow and Deep):
\begin{enumerate}
  \item 2-D lattice is flatten to a 1-D vector which is fed to the Input-layer. Output received from the NNs is also a vector which will be reconstructed into a 2-D Lattice.
  \item	Activation functions of each layer: Linear
   (other activations like $‘tanh’$ would have given a better result but we have tried to keep the models as simple as possible.)
  \item	Layers have no Biases.
  \item	L1 regularization is used to keep the model from over-fitting. L1 penalizes non-essential weights, thus encouraging their value to remain zero.
\end{enumerate}

Both shallow and deep nets undergo training with:
\begin{enumerate}
  \item	Loss Function: Hinge Loss
  \item	Optimizer: Adam (used to overcome the shortcomings of Gradient-descent.)
  \item	4000 epochs
  \item Batch size: 10
  \item Validation split: 20\% of the Training Dataset
\end{enumerate}

\subsection{\label{sec:inv_temp}Measuring the Inverse-Temperature($\beta$)}
\begin{figure}[h]
{\includegraphics[width=.5\linewidth]{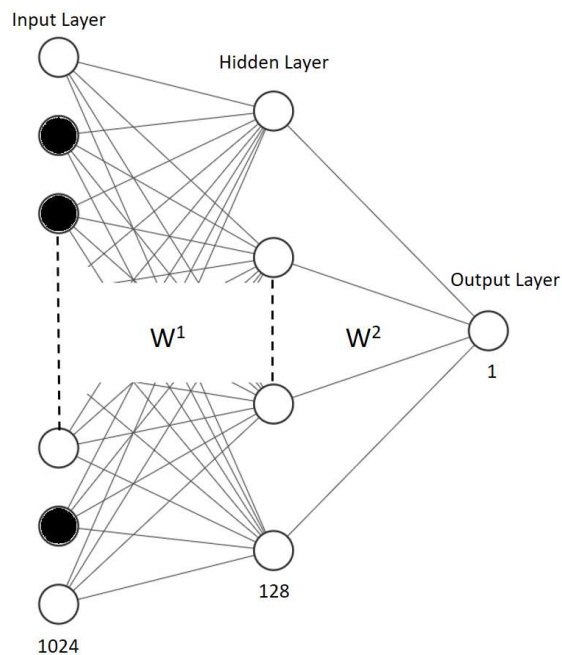}}
  \caption{\label{fig:NN_Temp} A 3-layered NN trained to predict inverse temperature $(\beta)$ using supervised learning.}
\end{figure}
Besides partition function(Z), temperature(T) and entropy(S) are some of the most important quantities that define an Ising lattice. Since we have been using inverse-temperature($\beta$) for our calculations we will designing a NNs to measure the inverse-temperature of the Ising lattice. For this we train a three-layer-deep NN to perform regression through supervised learning(Fig.\ref{fig:NN_Temp}).
The input layer ($Y^1$) receives the linearized spin configuration $(\sigma_{32 \times 32})$, hence the size of input layer is 1024 (Eq.\ref{eq:TL1}). The hidden layer ($Y^2$) consist of 512 neurons, with a `tanh' as the activation function (Eq.\ref{eq:TL2}). Finally the output layer ($Y^3$) consists of a single neuron which is used to predicted the inverse temperature of the input configuration, activation function used here is rectified linear unit or ReLU (Eq.\ref{eq:TL3}).
\begin{equation}
  \label{eq:TL1}
  Y_{i}^{(1)} = \text{Linearized } (\sigma_{m,n})
\end{equation}

\begin{equation}
  \label{eq:TL2}
  Y_{j}^{(2)} = \tanh\left(\sum_{i=1}^{1024} Y_{i}^{(1)}W_{i,j}^{(1)}+ b_{j}^{(1)}\right)
\end{equation}

\begin{equation}
  \label{eq:TL3}
  Y_1^{(3)} = \max \Bigg\{ 0,\left(\sum_{i=1}^{128} Y_{i}^{(2)}W_{i,1}^{(2)}+ b_{1}^{(2)}\right) \Bigg\}
\end{equation}
where $W_{i,j}$ is the weight connecting $i^{th}$ node of previous layer with $j^{th}$ the node of current layer and $b_{i}$ is the bias of $i^{th}$ node of the respective layer. This network will be trained with mean square err (MSE) as the loss function.

\section{\label{sec:Results}Results and Discussion}
\subsection{\label{sec:loss_learning}Losses and Learning}

\begin{figure}[ht]
{\includegraphics[width=.85 \linewidth]{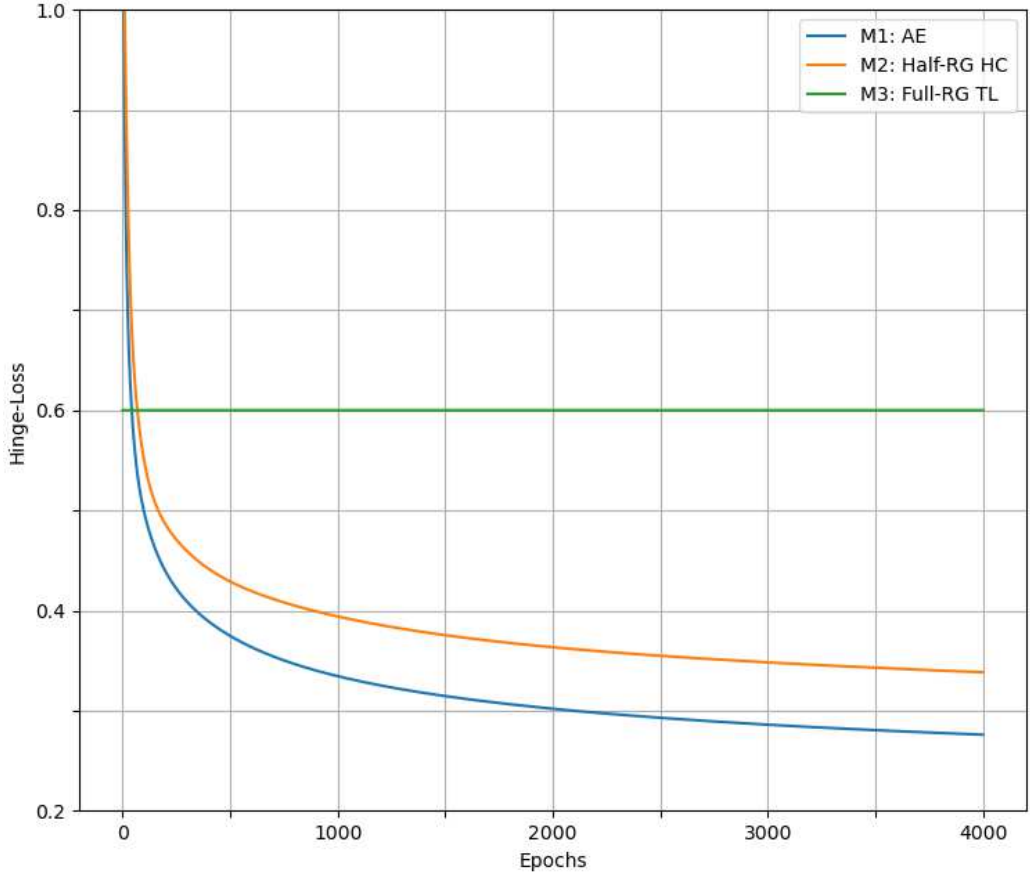}}
  \caption{\label{fig:training} Plot of training losses of Model-1(Blue) and Model-2(Orange) while training. And the final loss of Model-3(Green). See Fig. \ref{Fig:val_loss} in the appendix for validation losses.}
\end{figure}

\begin{figure}[ht]
{\includegraphics[width=.9\linewidth]{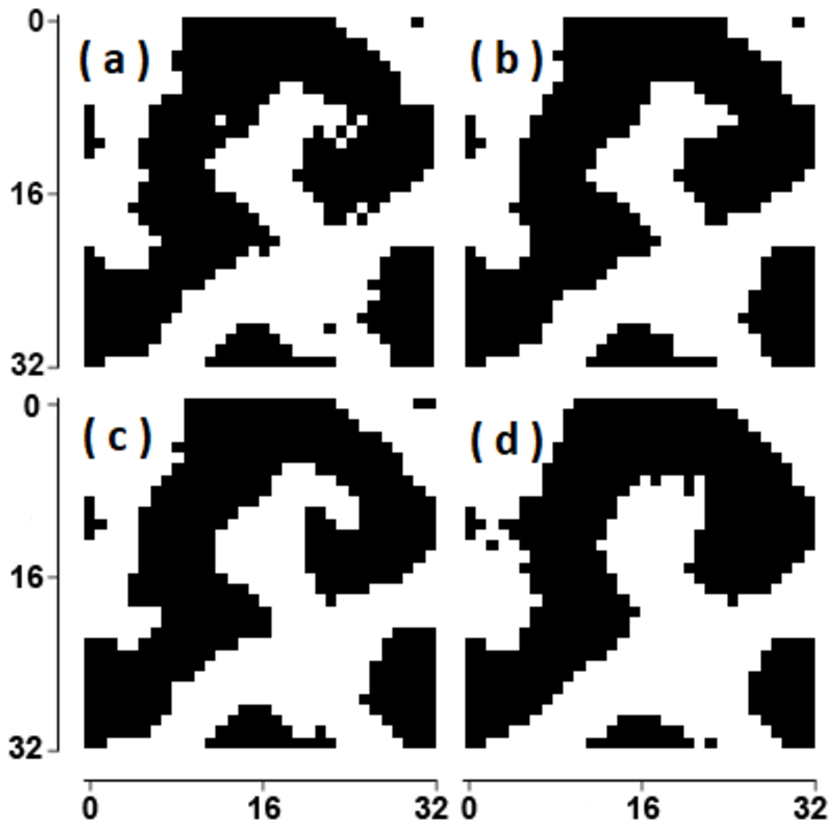}}

  \caption{\label{fig:lattice_compare} Reconstruction by various models (a) Input Ising Lattice $\parallel$  (b) Model-1: Autoencoder $\parallel$ (c) Model-2: RG Encoder-Auto Decoder $\parallel$ (d) Model-3: RG Encoder-RG Decoder.}
\end{figure}

After training the models for 4000 epoch, the hinge losses of both Models 1 and 2 are comparable 0.276 and 0.338 respectively, indicating that mapping of weights produced by RG is a type of encoder NN. The smaller loss of Model-1 with respect to Model-2 indicates that training through classic ML techniques such as back-propagation can provide much better encoding-decoding than the one provided by the RG (see Fig. \ref{fig:training}).\par
Model-3 is completely dependent on transfer learning for its training and we can run this NN across the Ising-dataset to check the loss it accumulates and the lattice configuration of the predicted output. This means that the relations learned via RG and Rescaling can act as an Encoder-Decoder.\par
However, Fig.\ref{fig:training} shows a substantial loss by Model-3 (0.6001) when compared to Model 1 and 2, even though its constituent shallow-nets have gone through the same number of epochs in training. The final validation losses of Model 1, 2 and 3 were 0.281, 0.341 and 0.601 respectively.  This indicates that the RG approach to learning is not efficient. Fig.\ref{fig:lattice_compare} shows a comparison of Ising lattice reconstructed by the three models. An increasing loss of information can be observed as we move from Autoencoder to Full-RG trained NNs. \par
This is probably due to the inherent difference that lies between these two approaches.

\textbf {Conjecture:} \emph{The goal of Renormalization group is to find a mathematically solvable framework for coarse-graining, whereas the job of Autoencoder(and Deep-Learning is general) is to find a mapping with minimum loss of information.}\par
Model-3 can also be modified further with hard-coded weights, which carry the exact-mapping used in real space RG techniques like decimation/majority/naive coarse-graining.

\subsection{\label{sec:weights}Weights and Spin-Correlation matrices}

\begin{figure}[ht]
{\includegraphics[width=.99\linewidth]{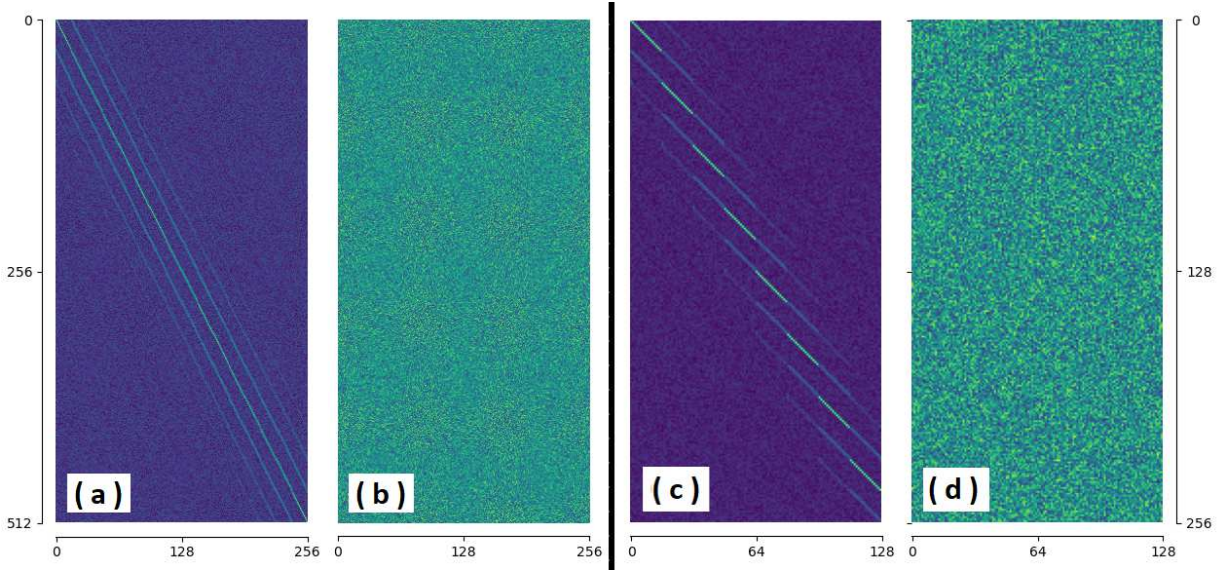}}
  \caption{\label{fig:weights1} Comparison of weight-matrix ($W_1$) $1024\times512$ connecting layer ($I$) with ($H_1$).  (a) Trained on RG $\parallel$ (b) Trained on AE. As well as weight-matrix ($W_2$) $512\times256$ connecting layer $H_1$ with $H_2$.  (c) Trained on RG $\parallel$ (d) Trained on AE.}
\end{figure}

Analysis of weights(W) acquired through training on all 3 models shows that ML and RG follow very different part to optimization.
DL is essentially a careful calibration of weights (and biases) which minimizes the loss function, therefore it would be fruitful for us to visualize the weights acquired by the NNs as they complete their training. A plot of values aquired by weights between various layers of NNs are shown in the Fig.\ref{fig:weights1}. As we can see the layers of Model 2 have a very sparse configuration (Fig.\ref{fig:weights1}(a), \ref{fig:weights1}(c)), while the weights in the layers of Model 1 (the Auto-Encoder) are more homogenous (Fig.\ref{fig:weights1}(b), \ref{fig:weights1}(d)). This shows a stark difference between learning acquired through RG and Autoencoders. The pattern we observe in RG trained weight matrices can explained through Eq. \ref{eq:w1} and \ref{eq:w2}, i.e. the trained weights are trying to mimic the mapping done through hard-coding in Sec. \ref{sec:hard-code}. \par

\begin{figure}
{\includegraphics[width=0.9\linewidth]{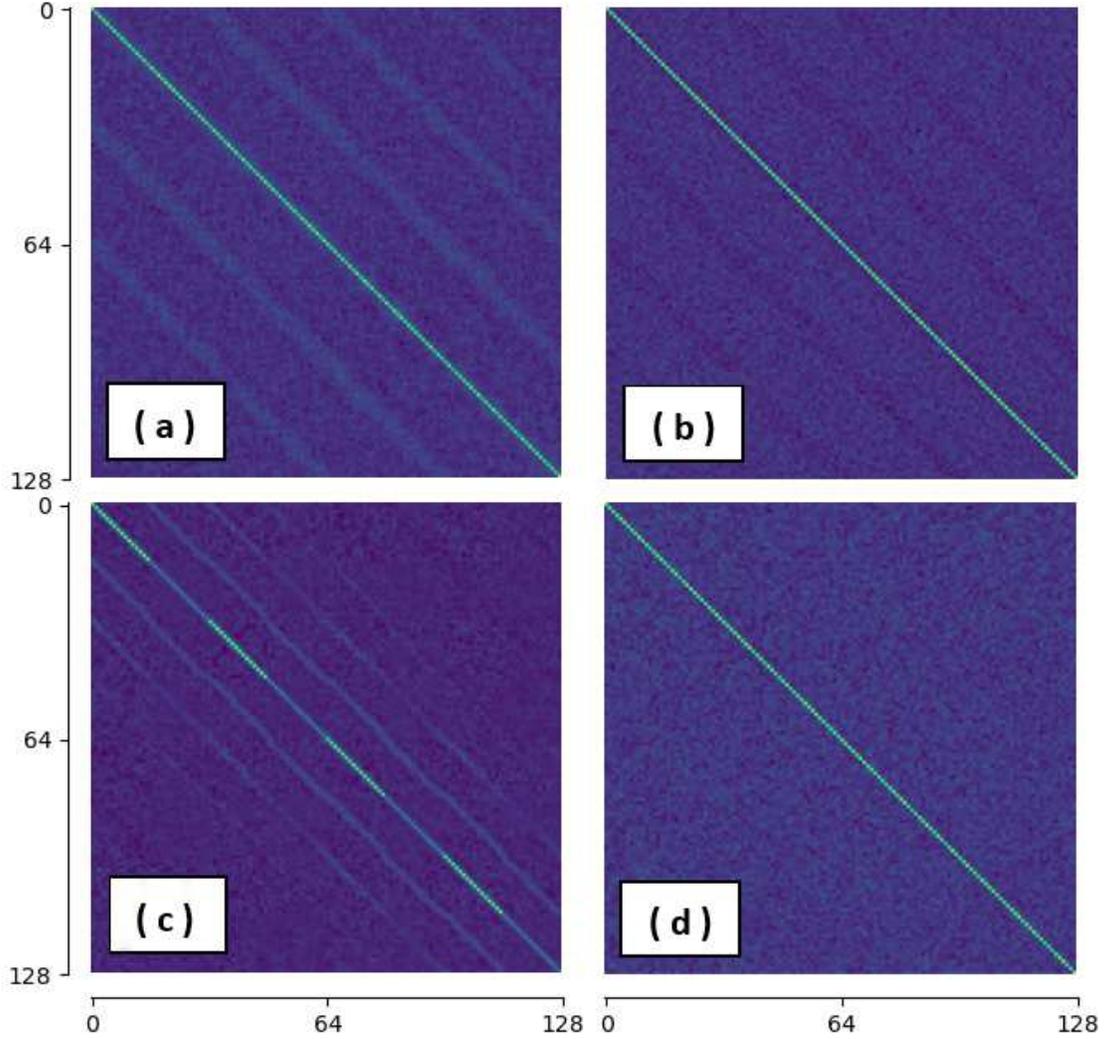}}
  \caption{\label{fig:WWT} Comparison of spin-correlation matrix ($WW^T$): $1024\times1024 (W_1W_1^T)$ (a)  Trained on RG  (b) Trained on AE. $\parallel$ $512\times512 (W_2W_2^T)$ (c)  Trained on RG (d) Trained on AE.}
\end{figure}

Another important result for comparison is the spin correlation matrix $WW^T$, which helps us in understanding the relations formed within the NNs \cite{Iso2018}. Although our models were trained differently than the Restricted Boltzmann Machines(RBMs) used by \textcite{Iso2018}, values in $WW^T$ will carry similar significance. The heat-map of $WW^T$ matrix from encoding layers of model 1 and 2 are shown in (Fig.\ref{fig:WWT}). As we can see, large values are mostly concentrated on the diagonal elements $(\sigma'_{i,j})$ where $i=j$, justifying the self-correlation of the input spins. However, presence of large valued near-diagonal elements($j = i \pm 32, 16 $) in model-2 signify that the layers with RG mapping has more spins-correlation than the layers of Autoencoders(model-1). Please recall that the lattice($32 \times 32$) was linearized for input, therefore spin$(\sigma_{i,j})$ where $(i,j=0,...,31)$ will be transformed to $\sigma^L_{k}$ where $(k=0,...,1023)$. As shown by \textcite{Iso2018} this might signify that the AE has learned much more non-relevant information from the dataset than necessary. This indicates over-fitting, with relations `diffused' throughout the lattice. Indicating that an AE with smaller hidden layers may perform more efficiently than the current one.\par
NOTE: The figures shown here are zoomed-in versions of the actual matrices, please see Fig.\ref{hi-res:W1} for $W_1$ and Fig.\ref{hi-res:W2} for $W_2$. For spin correlation matrices see Fig.\ref{hi-res:WWT1}, \ref{hi-res:WWT2} in the appendix.

\subsection{\label{sec:entropy}Entropy and Compression}

\begin{figure}[ht]
{\includegraphics[width=0.8\linewidth]{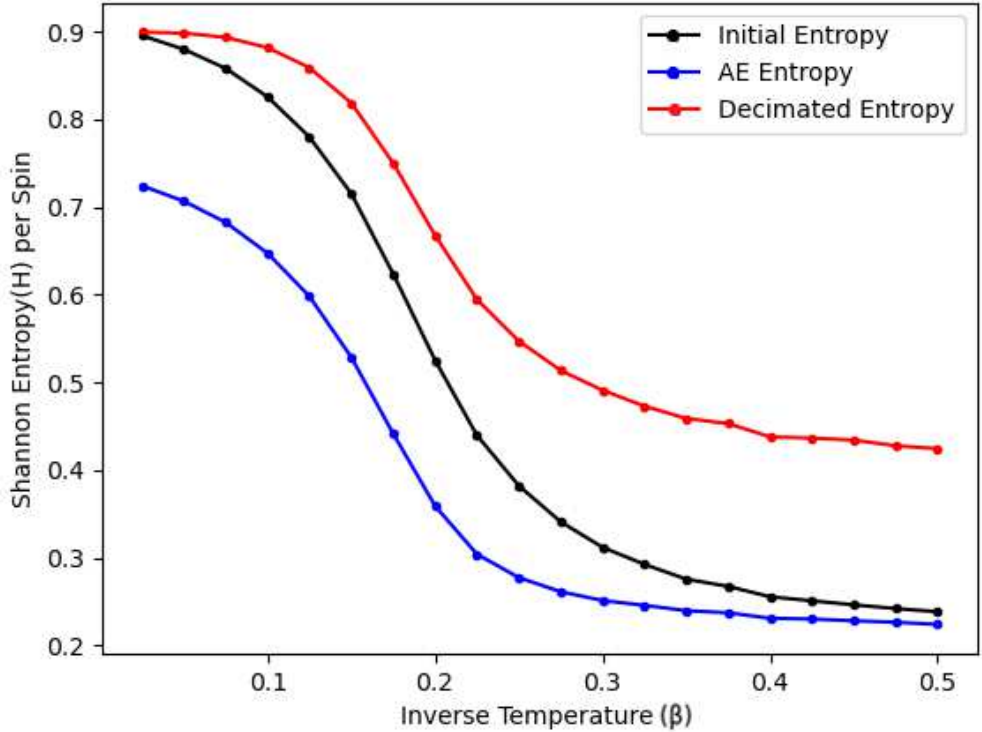}}
  \caption{\label{fig:entropy}Plot comparing Shannon-Entropy of the initial Ising lattice with output lattices from Autoencoder and RG decimation.}
\end{figure}

As is the case in statistical mechanics, Entropy and its interpretations hold a lot of significance information theory(IT). Hence Shannon's entropy \cite{shannon1948} (as it's called in IT) signifies the actual amount of information the data carries. Since coarse-graining done by RG or dimensionality reduction done by AE are essentially data-compression, calculating and comparing the entropy of the lattices generated by these models will give us an interesting insight into their compressing power. All this can be done with the help of well-established methods in Image-processing. Since we are going to study the change in entropy over a range of $\beta$, it would be prudent to retrain a NN of Model-1(AE) configuration on the 50,000 spin configurations generated for $ \beta \in [0.1,0.6)$, lets call it AE-wide-range(AE-WR), see Fig.\ref{Fig:train_val_loss} for training losses. Decimation is not dependent on $\beta$, hence require no changes.

\begin{figure}[ht]
{\includegraphics[width=0.8\linewidth]{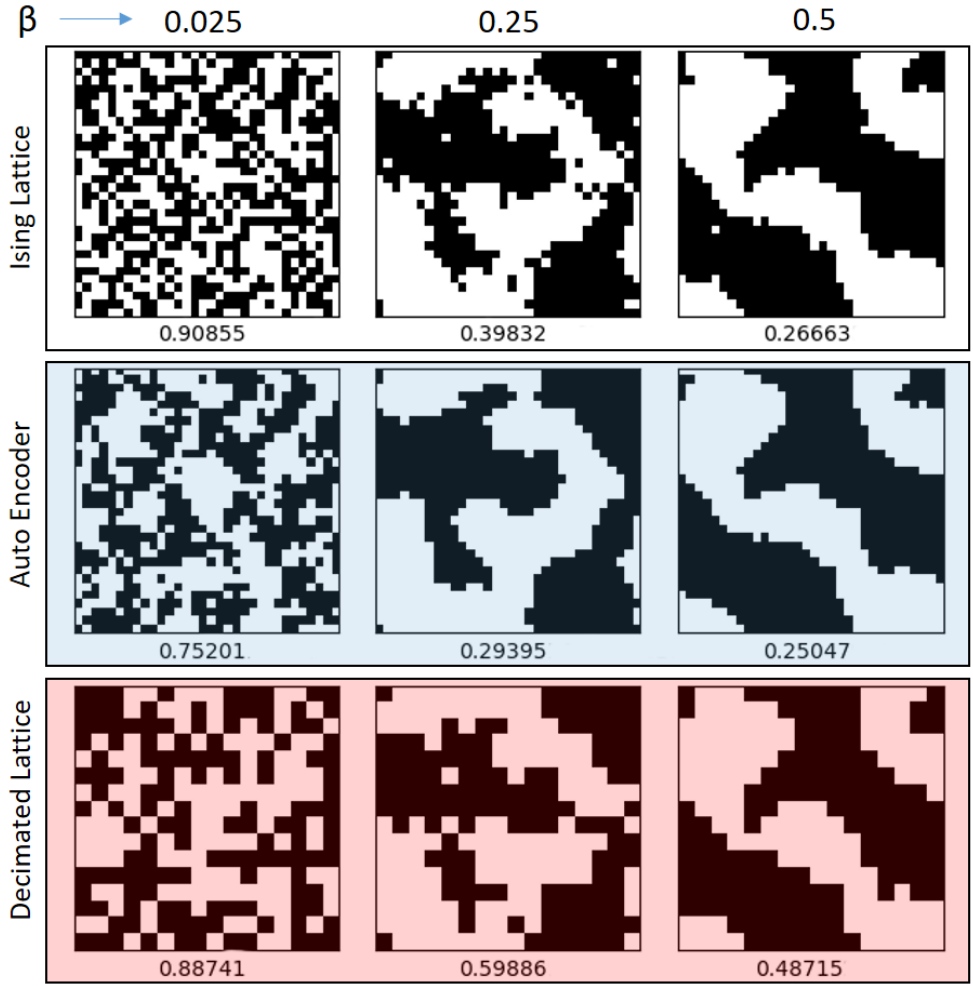}}
  \caption{\label{fig:entropy_lat} Comparing the output lattices of: ($1^{st}$ row) Ising lattice $32\times32$ at $\beta = 0.025, 0.25, 0.5$ respectively. $\parallel$ ($2^{nd}$ row) Autoencoder $32\times32$. $\parallel$ ($3^{rd}$ row) RG decimation $16\times16$.}
\end{figure}

In our case, we are interested in calculating the entropy in the neighbourhood of each spin in the Ising lattice. For this, will we take the help of Scikit-Image processing library \cite{Ent_skimage}. We have generated a thousand lattices for each of the 20 values of $\beta$ between 0.025 and 0.5. Each lattice also goes through the autoencoder and coarse-graining which generates two additional sets of lattices. Entropy is calculated in $3\times3$ neighbourhood each spin of the lattice, which is then averaged to the value of local entropy per spin of a single lattice. This process is repeated for all 1000 lattices and the average Entropy is calculated at the particular $\beta$ for all 3 sets.\par

The plot in Fig.\ref{fig:entropy} indicates that at lower $\beta$ (high temperature, high entropy), output from AE has much less entropy as compared to the original lattice indicating the compression done by AE is quite lossy in nature, as it tends to smoothen out the insignificant degrees of freedom (Fig.\ref{fig:entropy_lat}). At higher $\beta$ (low temperature, low entropy) however, the output entropy from AE converges with the original lattice while the decimated lattice from RG retains some of the smaller features, increasing its entropy(relatively). As we can see from the $2^{nd}$ row of Fig.\ref{fig:entropy_lat} the autoencoder has smoothened the features of the original lattice. This `de-noising' property of AE is well known \cite{vincent2008}. Decimation on the other hand increases the randomness around the neighbourhood of the spin, retaining some extra entropy($3^{rd}$ row).
\begin{figure}[ht]
{\includegraphics[width=0.69\linewidth]{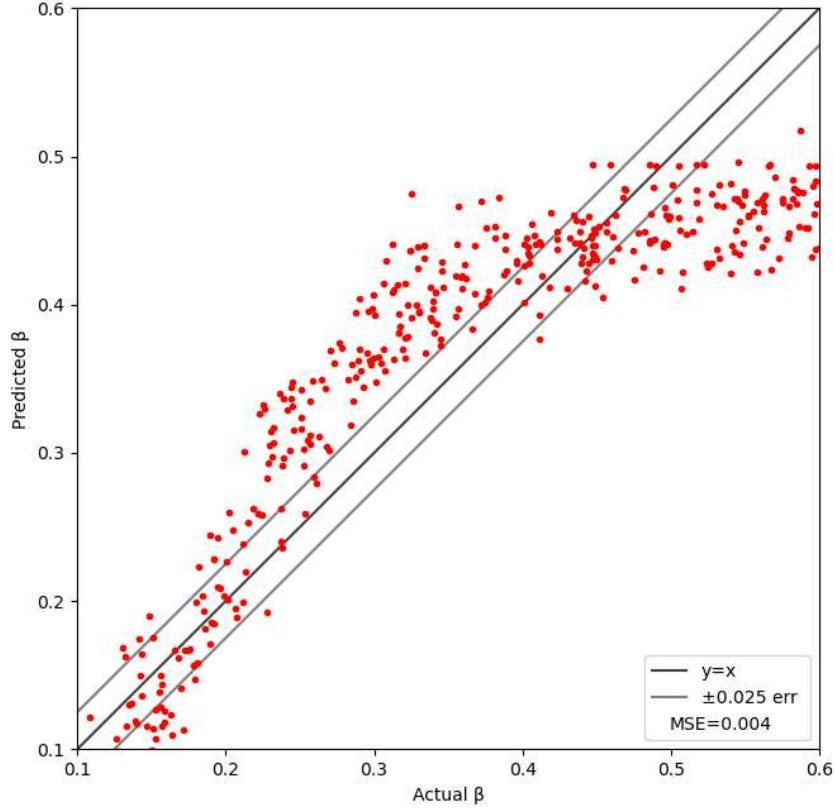}}
  \caption{\label{fig:beta_pred} Plot of the Actual-$\beta$ vs Predicted-$\beta$ produced by NN trained to measure the inverse temperature of the given Ising lattice.}
\end{figure}

\begin{figure}[ht]
{\includegraphics[width=0.7\linewidth]{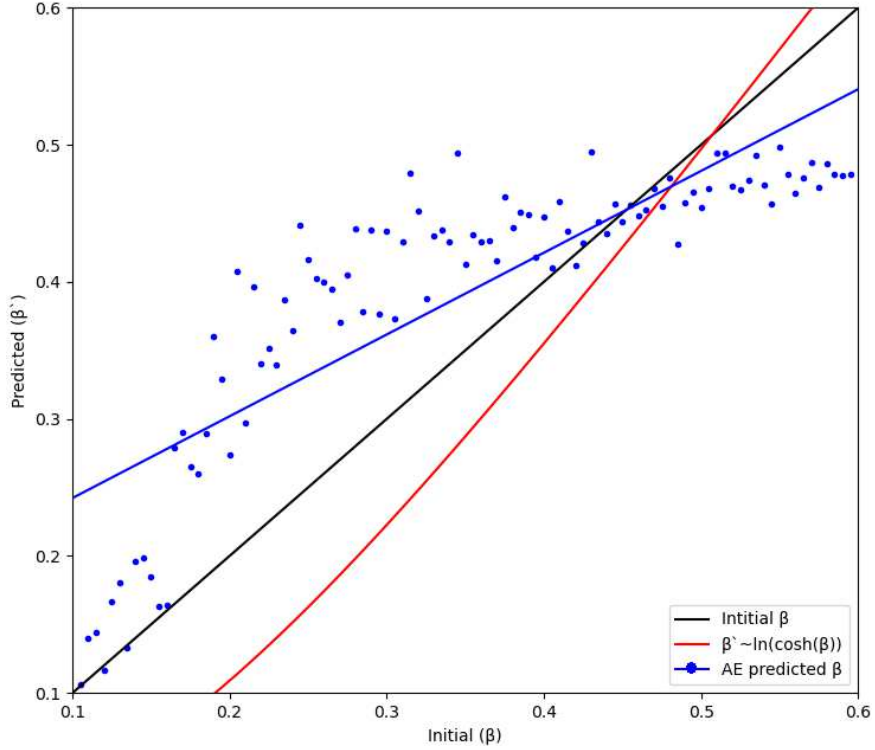}}
  \caption{\label{fig:beta_compare} Plot showing the $\beta$ of lattices produced by AE compared with the mathematical results from RG decimation.}
\end{figure}

\subsection{\label{sec:RG_flow}Inverse Temperature and the RG Flow}

The $\beta$-predicting NN described in Sec.\ref{sec:inv_temp} was trained on Ising lattices for $ \beta \in [0.1,0.6)$  until the mean square err(MSE) came down to 0.0035, then a new set of 10,000 lattices were used to check the test-losses of the NN (see Fig.\ref{fig:beta_pred}). This NN was then deployed to predict the $\beta$ of Ising lattices produced by the AE-WR and was compared with the mathematical results for RG decimation shown in Sec.\ref{sec:RG_deci}.\par
As we have seen in Eq.\ref{eq:P_final}, the new $\beta'\sim\ln(\cosh(\beta))$ after every decimation, this causes $\beta$ to diverge from it's original value, remaining unchanged only at fixed point(s) like Critical Beta($\beta_c$) where $\beta'=\beta$. This divergence/convergence from a fixed point is called RG flow, which will help us in exploring some more the differences between AE and RG.\par

As discussed in the previous section, Autoencoders have a de-noising property \cite{vincent2008}, which reduces the randomness in the spin configuration, increasing the $\beta'$ of the system, hence the system with $\beta < \beta_c$ flows toward $\beta_c$. At $\beta > \beta_c$ however, AE reduces the $\beta$ back to $\beta_c$, again flowing towards the fixed point. This indicates a stark difference in mechanism of learning done by AE and coarse graining of RG (Fig.\ref{fig:beta_compare}).\par

It also worth noting that the Blue-line fitting the Predicted-$\beta$ from AE intersects the $\beta=\beta'$ line (Black) at $\beta = 0.45265$ which is much closer to the actual fixed-point $\beta_c=0.4352$ than the approximation done in Sec.\ref{sec:RG_deci} for RG(Red) at $\beta=0.505$.

\subsection{\label{sec:Conclusion}Discussions}
Our investigation into the parallels between RG and DL has lead us to some anticipated results and some surprising revelations. Decimating did turn out to be a viable scheme on which deep-net could be trained to encode Ising lattices. However, it did not seem to be as efficient as the classic DL techniques, with a substantial difference between the losses RG accumulated even when the hardcoded schema was deployed. We can attribute this disparity to the objectives behind doing RG which is very different from objectives assigned to AE. We also studied the internal structures formed inside both techniques and discovered that they were quite different.  While RG formed a sparse pattern, the mapping that emerged from AE was homogenised and diffused throughout the network. Moreover, the nodes/spins in the AE schema were weakly correlated with each other as compared to RG, indicating that AE can be further increased. This means AE is more efficient than the RG as far as the number of network nodes was concerned. \par
Since both RG and DL are performing dimensionality reduction, studying the compression ratio and information loss across the range $\beta$ becomes the next logical step. As entropy and information are parallel interpretations of the same quantity, we analysed the entropy of the outputs from RG and AE, then compared them with the entropy of the input Ising lattice. The $\beta$-dependence of information loss due to compression revealed a clear distinction between the two processes. While entropy of the input lattice was comparable to RG lattice at low $\beta$, it converged to the entropy of the AE lattice at higher $\beta$.\par
Finally, we shifted our focus to RG flow and set out to find anything similar happening in deep learning. For this, we utilized a separate NN trained to predict the inverse temperature($\beta$) of the input lattice. It was then employed to detect the change in $\beta$ of the output from the AE and compared it to the mathematically derived result for the RG. We found that the outputs from AE indeed show an RG-like flow with $\beta_c$ as its fixed point. Surprisingly, however, the direction of flow was opposite. Where RG flowed away from the $\beta_c$, DL flowed towards it. Essentially changing $\beta_c$ from a repeller to an attractor, which appears contrary to Wilson's interpretation of `relevant' couplings/operators in RG \cite{Wilson1972, Wilson1983}. Further exploration in this area may help us in establishing a correspondence between renormalization, machine learning and epistemic learning. In our view, the arrival of the DL algorithm at the fixed point should be seen as `learning’ about the criticality of the Ising model, while coarse-graining the system leads to `unlearning’ of the information about the lattice. In other words, with auto-encoding, the unsupervised algorithm will learn about the criticality of the 2D Ising model, knowing nothing about the mathematics governing the model beforehand. On the other hand, with decimation, we are progressively losing knowledge about the criticality and mechanics of the Ising model. In this sense, RG through the decimation of 2D Ising lattice can hardly be called `learning'. We suspect the same can be said for other divergent RG flows.

\begin{acknowledgments}
We would like to thank Ministry of Education, Govt. of India for funding the research through Indian Institute of Technology Patna. Figures showing the architecture of the neural networks were made using: LeNail (2019), NN-SVG: Publication-Ready Neural Network Architecture Schematics;
Journal of Open Source Software, 4(33), 747, https://doi.org/10.21105/joss.00747
\end{acknowledgments}



\bibliography{RG-AE_Shukla_et_al_211129.bib}

\newpage

\appendix

\setcounter{figure}{0}

\section{\label{sec:train_plots}Training Plots}

\makeatletter
\renewcommand{\thefigure}{A.\@arabic\c@figure}
\makeatother

\begin{figure}[bh]
\includegraphics[width=\linewidth]{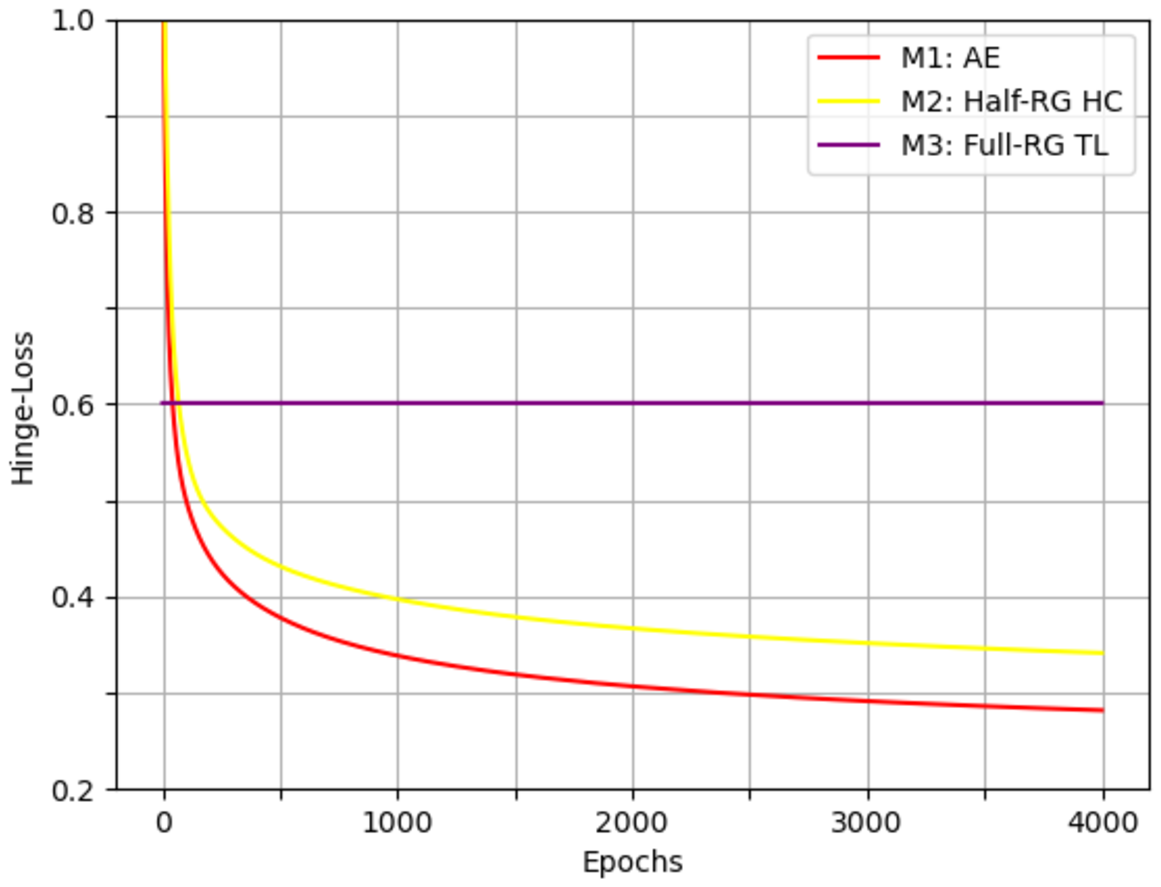}
  \caption{\label{Fig:val_loss} Plot of validation losses of Model-1(Red) and Model-2(Yellow) while training. And the final loss of Model-3(Purple). Validation losses after 4000 epochs are 0.281, 0.341 and 0.601 respectively. Corresponding training losses were 0.275, 0.338 and 0.6 respectively.}
\end{figure}

\begin{figure}[th]
\includegraphics[width=0.9\linewidth]{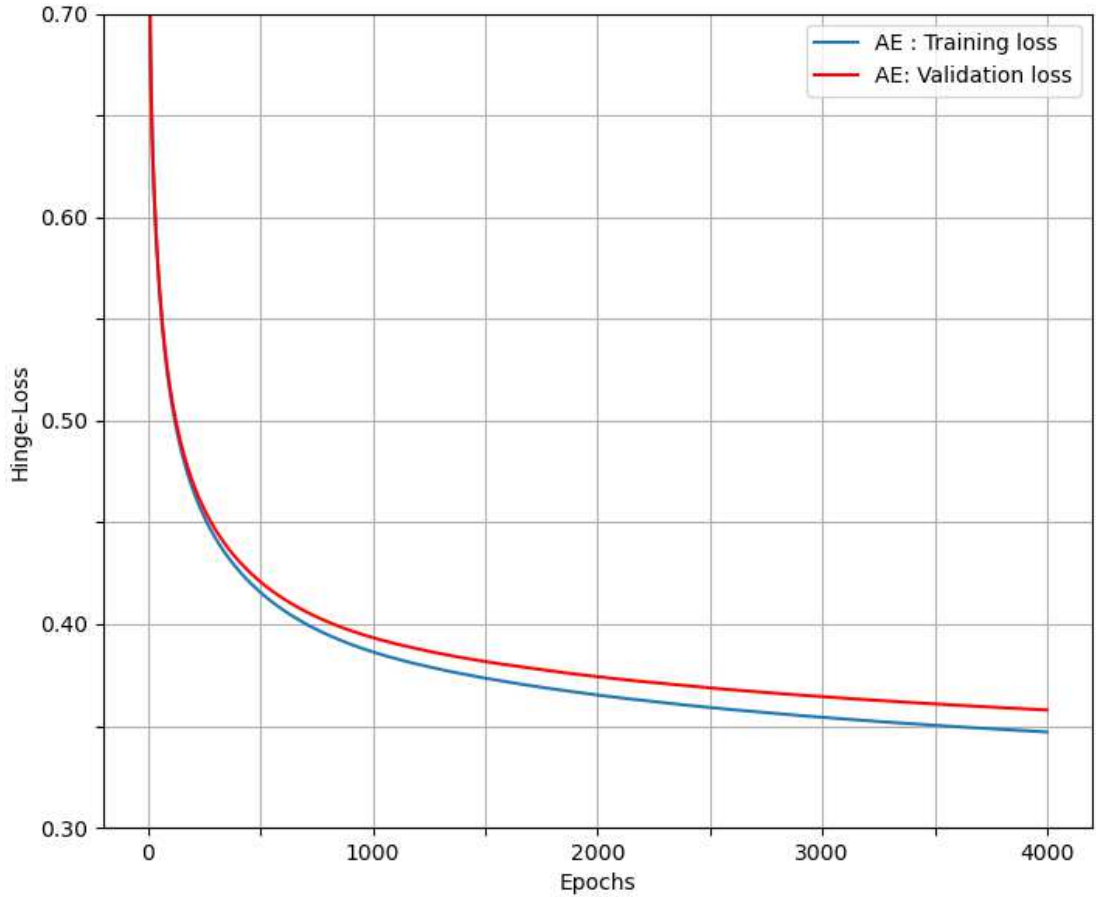}
  \caption{\label{Fig:train_val_loss} Plot of training(Blue) and validation(Red) losses of AE-WR while training. Final validation loss after 4000 epochs was 0.357, training loss was 0.347.}
\end{figure}

\newpage

\setcounter{figure}{0}

\section{\label{sec:Hi-res}High Resolution Images}
Here are the High-Resolution Images mentioned in the previous sections.

\makeatletter
\renewcommand{\thefigure}{B.\@arabic\c@figure}
\makeatother

\begin{figure}
\includegraphics[width=0.42\linewidth]{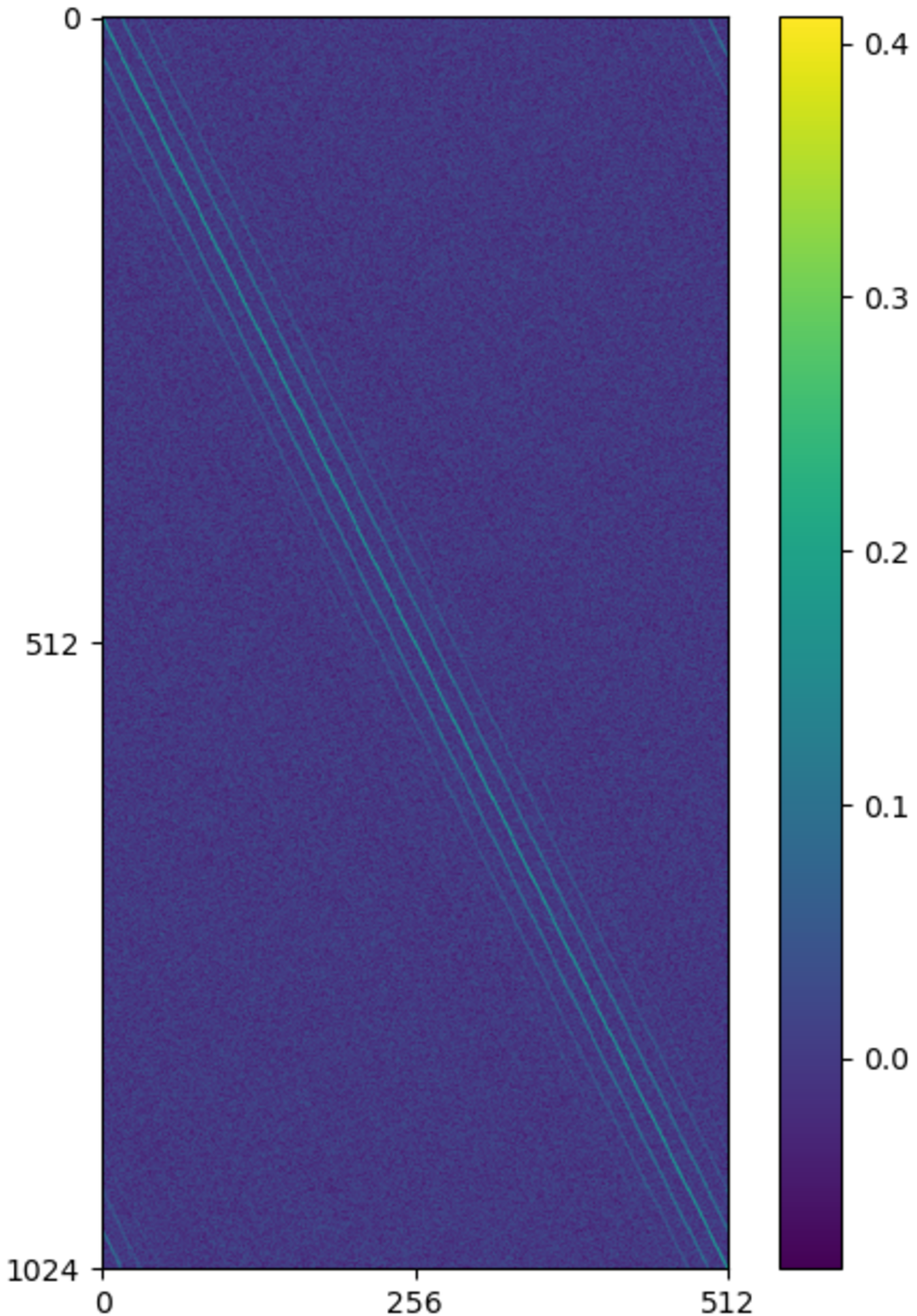}
\includegraphics[width=0.439\linewidth]{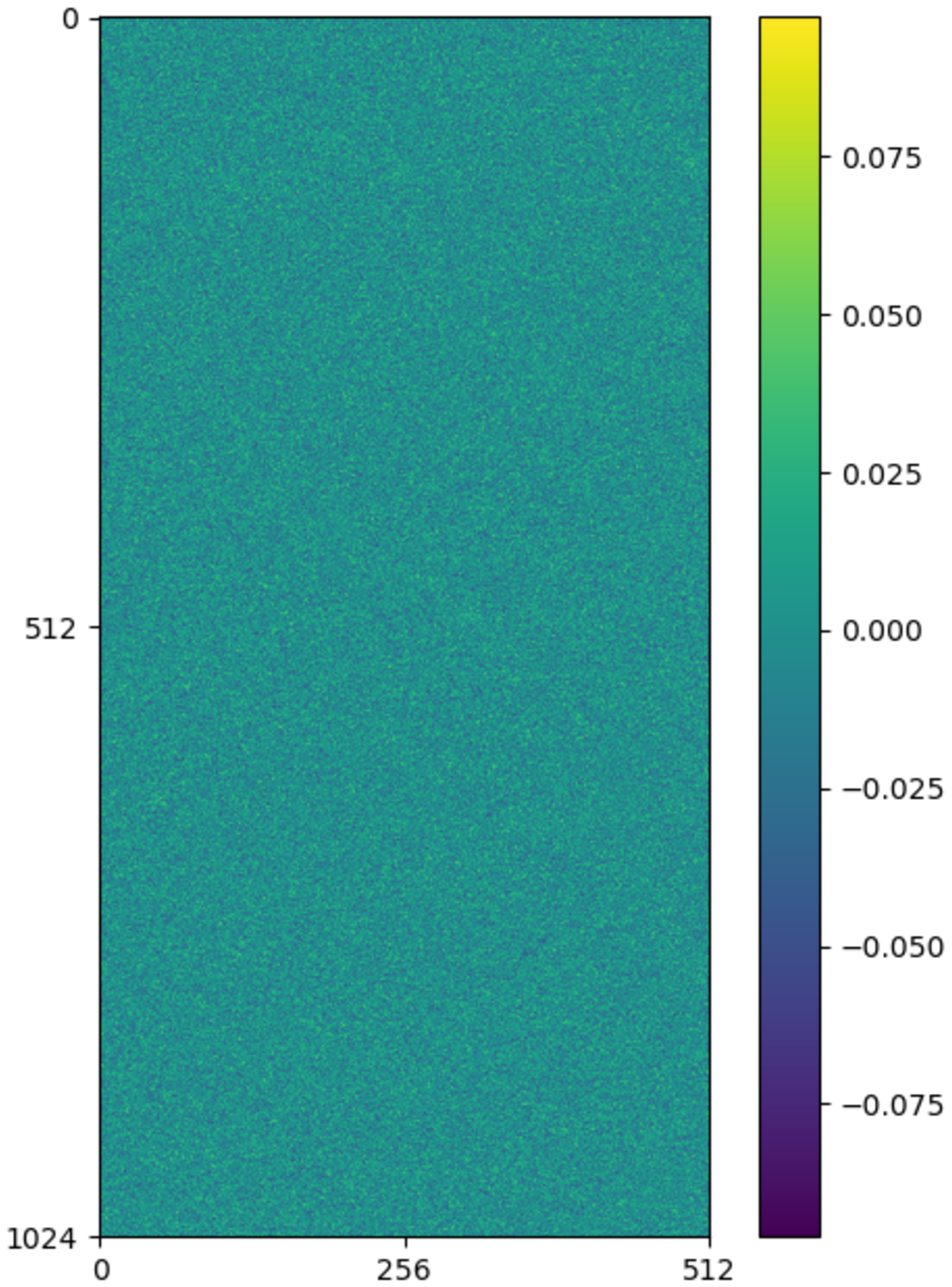}
  \caption{\label{hi-res:W1} Weight-matrix ($W_1$) $1025\times512$ connecting layers $I$, $H_1$.  (Left) RG $\parallel$ (Right) AE.}
\end{figure}

\begin{figure}
\includegraphics[width=0.42\linewidth]{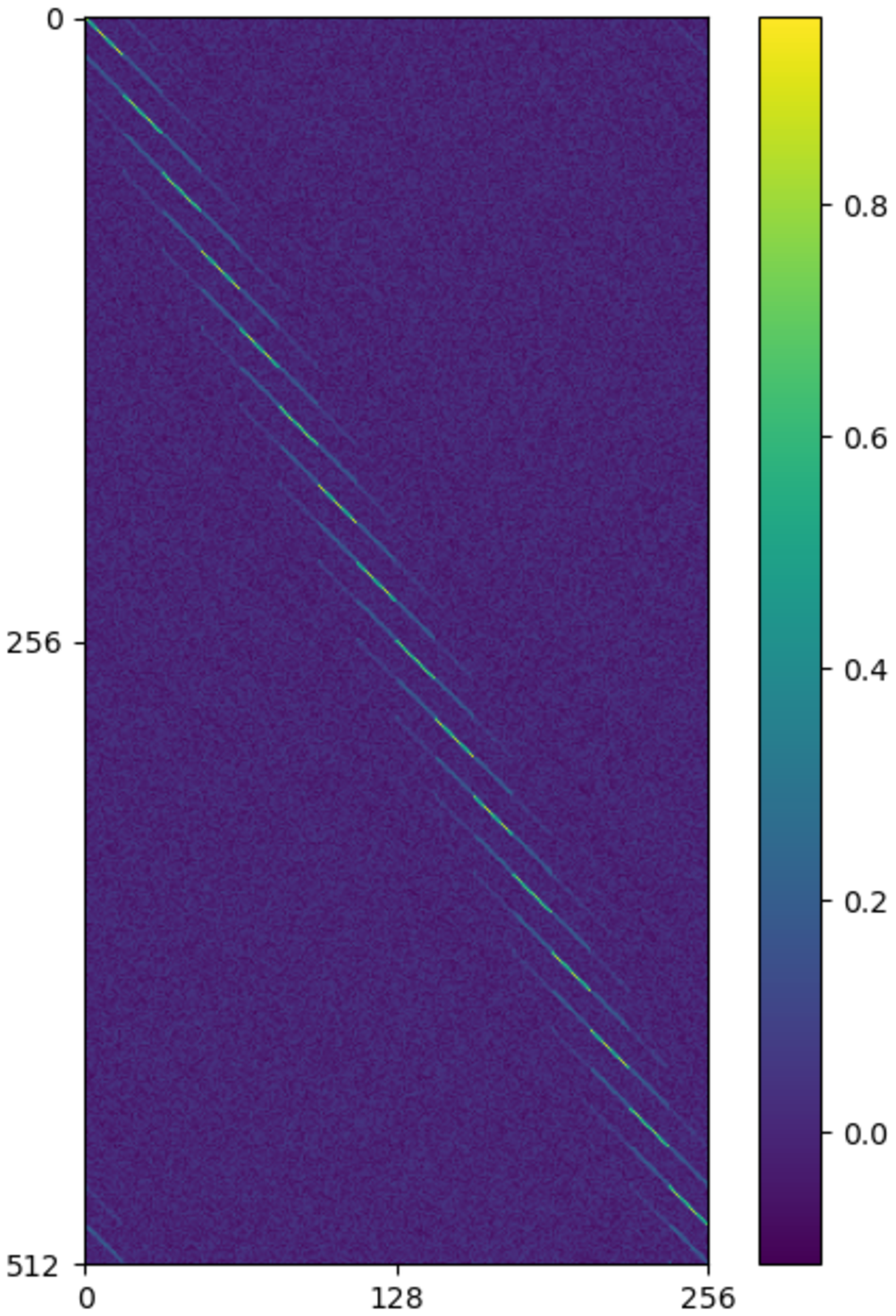}
\includegraphics[width=0.439\linewidth]{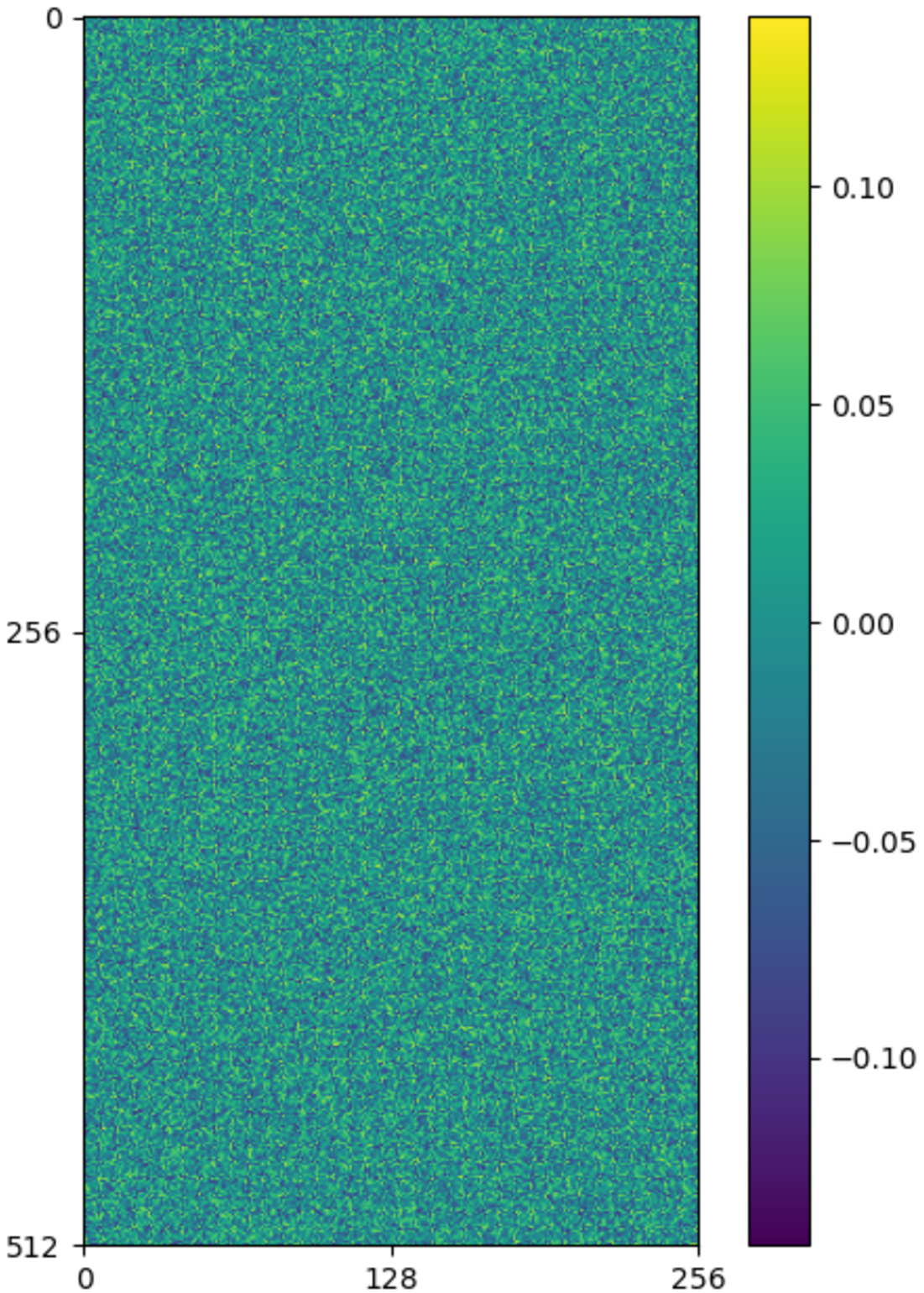}
\caption{\label{hi-res:W2} Weight-matrix ($W_2$) $512\times256$ connecting layers $H_1$, $H_2$. (Left) RG $\parallel$ (Right) AE.}
\end{figure}.

\begin{figure}
\includegraphics[width=0.482\linewidth]{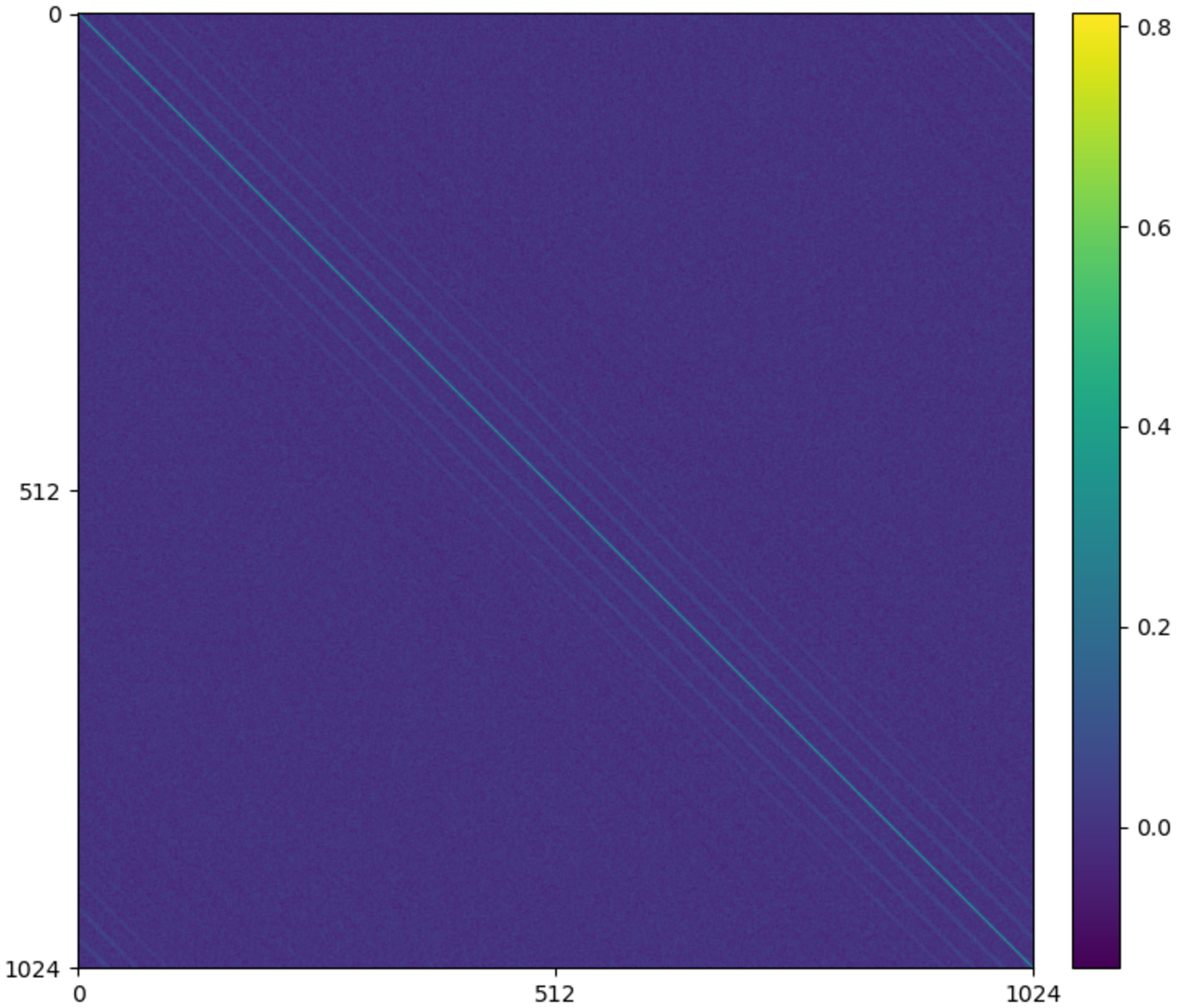}
\includegraphics[width=0.487\linewidth]{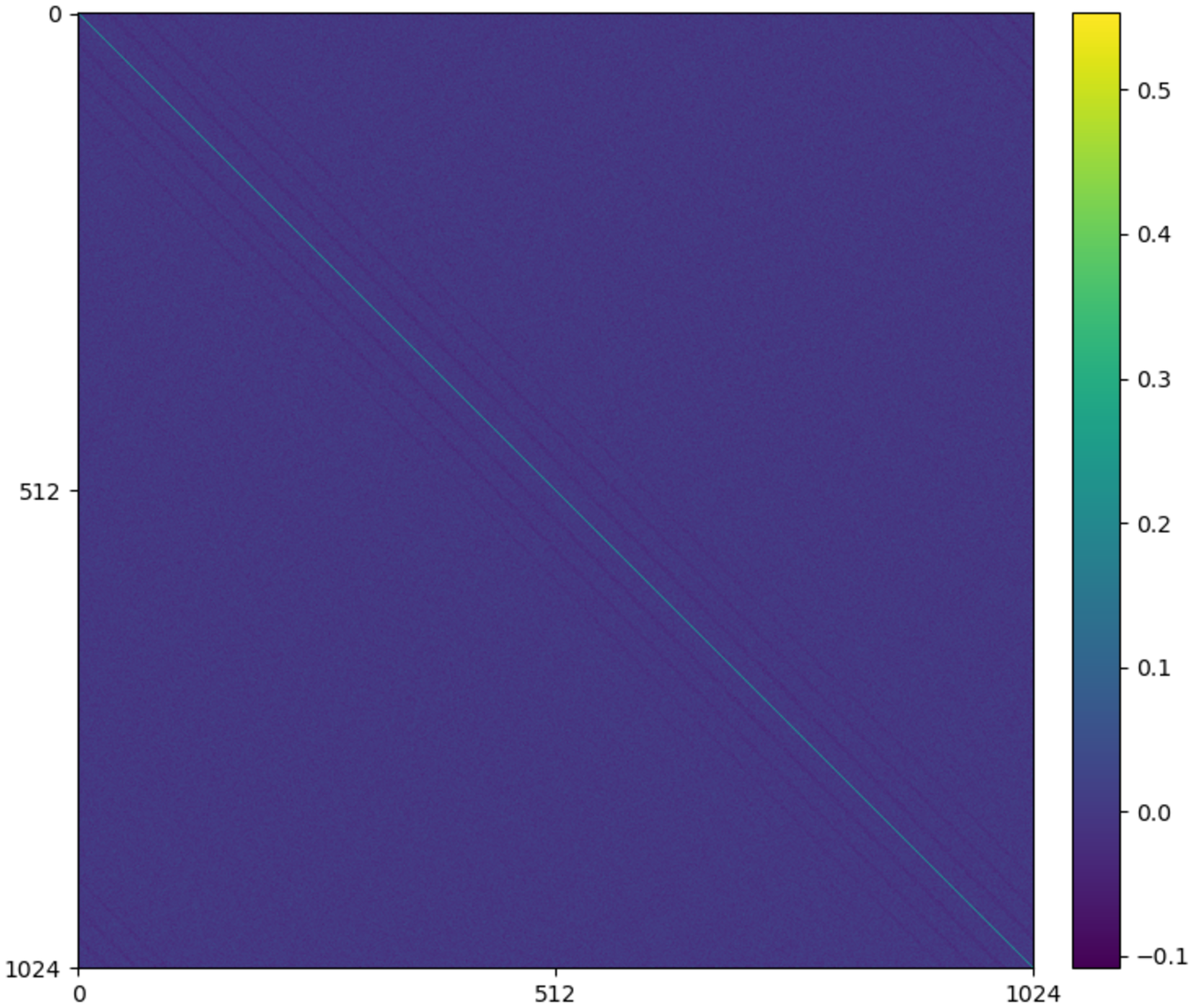}
  \caption{\label{hi-res:WWT1} Spin-correlation matrix $W_1W_1^T$ $(1024\times1024)$: (Left)  Trained on RG $\parallel$ (Right) Trained on AE.}
\end{figure}

\begin{figure}
\includegraphics[width=0.482\linewidth]{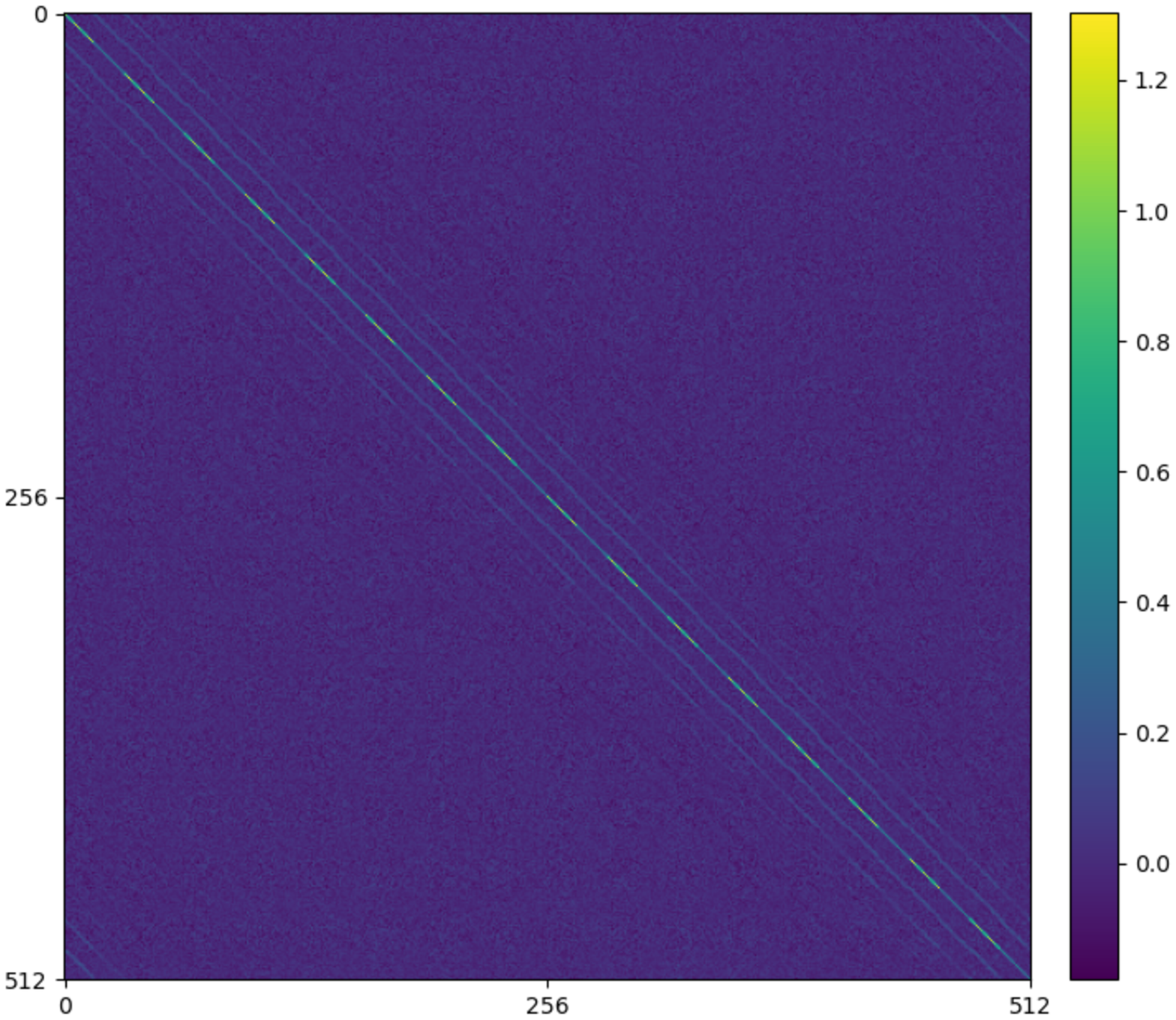}
\includegraphics[width=0.487\linewidth]{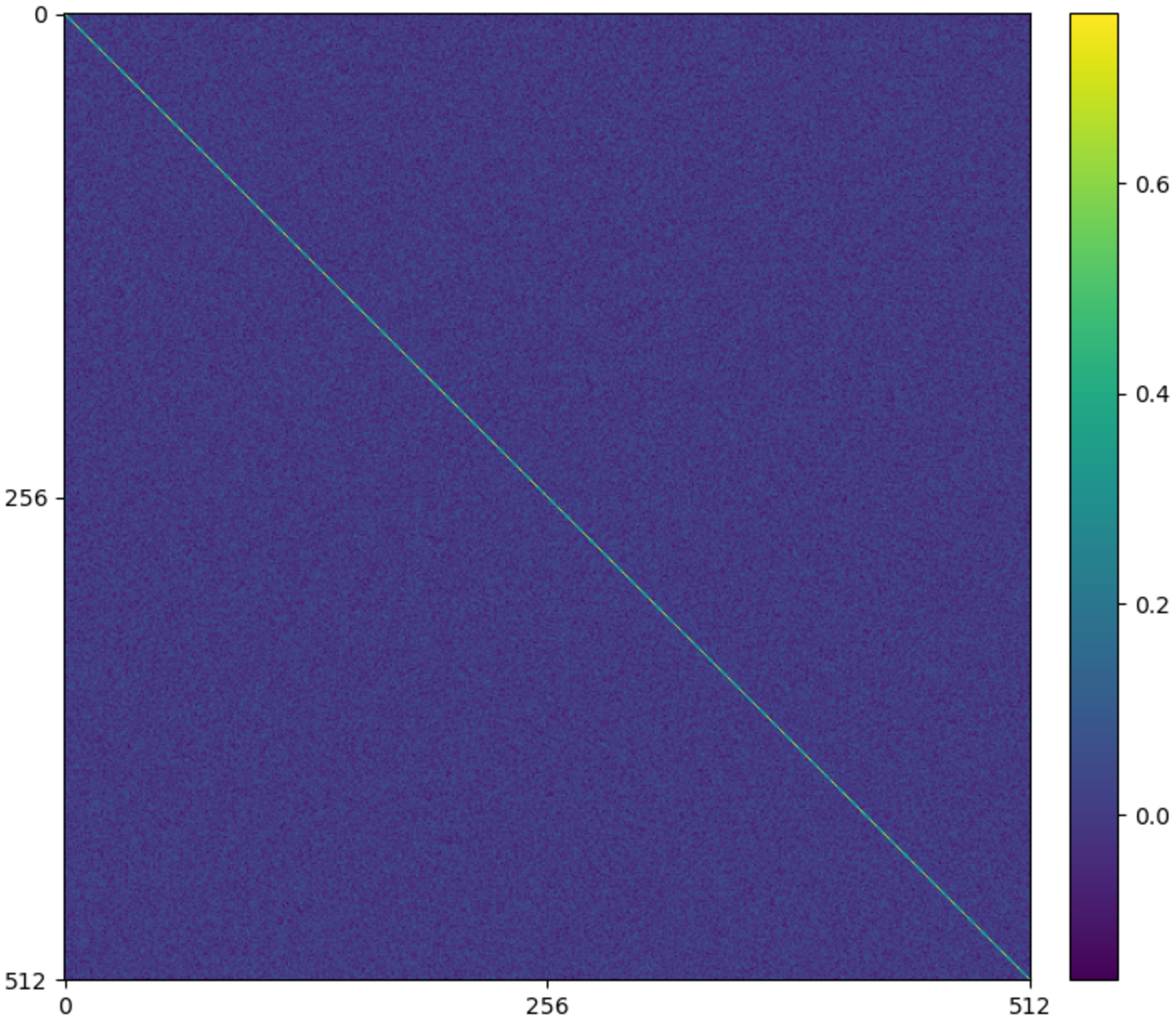}
  \caption{\label{hi-res:WWT2} Spin-correlation matrix $W_2W_2^T$ $(512\times512)$: (Left)  Trained on RG $\parallel$ (Right) Trained on AE.}
\end{figure}

\clearpage

\section{\label{sec:Program}Program Skeleton}
\begin{enumerate}
  \item Include libraries;
  \item Initialize hyper-parameters;\par
  ...
  \item Initialize a Multilayer Perceptron;
  \item Using a tree based model, add layers;
  \item Compile the Model with optimizer, loss function;\par
  ...
  \item Load the Dataset;
  \item Parse the data-files into variable tensors;
  \item Normalise and resize tensors;
  \item Load weights from external source if using Transfer-learning;\par
  ...
  \item Fit the Model into the training data with ample epochs;
  \item Test the accuracy of the trained model on the Testing-Dataset;\par
  ...
  \item Plot the graph of Loss Function vs Epochs;
  \item Plot the actual Output vs predicted Output
  \item Save the trained Model, weights and biases:
  \item End;

\end{enumerate}

\section{\label{sec:table}Table comparing models}
\begin{table}[ht]
  \centering
  {
  \begin{tabular}{ |p{3cm}||p{4.5cm}|p{4.5cm}|p{4.5cm}|  }
   \hline
   \multicolumn{4}{|c|}{ Comparison of the 3 models} \\
   \hline
    &\thead{Model-1 \\Autoencoder} & \thead{Model-2 \\TL Encoder \\Unsupervised Decoder} & \thead{Model-3 \\TL Encoder \\TL Decoder}\\
   \hline
   \hline
   \thead{Input-layer: ‘I’\\ Size: 1024}   & \thead{Input  of a linearized\\ Ising Lattice of\\ 32x32}  & \thead{Input  of a linearized\\ Ising Lattice of\\ 32x32 } &\thead{Input  of a linearized\\ Ising Lattice of\\ 32x32}\\
   \hline
   \thead{Hidden-layer: ‘H1’\\ Size: 512} &\thead{Encode input from ‘I’\\ down to 512 units.\\ Trainable.}  &\thead{Encode input from ‘I’\\ down to 512 units.\\ Not trainable as\\ learning is transferred from \\HCG-1024-512}    &\thead{Encode input from ‘I’\\ down to 512 units.\\ Not trainable as\\ learning is transferred from \\CG-1024-512}\\
   \hline
   \thead{Hidden-layer: ‘H2’\\ Size: 256} &\thead{Encode input from ‘E1’\\ down to 256 units.\\Trainable.} &\thead{Encode input from ‘E1’\\ down to 256 units.\\ Not trainable as\\ learning is transferred from \\HCG-512-256} &\thead{Encode input from ‘E1’\\ down to 256 units.\\ Not trainable as\\ learning is transferred from\\ CG-512-256}\\
   \hline
   \thead{Hidden-layer: ‘H3’\\ Size: 512} &\thead{Decode input from ‘E2’\\ up scaled to 512 units.\\ Trainable.} &\thead{Decode input from ‘E2’\\ up scaled to 512 units.\\ Trainable.} &\thead{Decode input from ‘E2’\\ up-scaled to 512 units.\\ Not trainable as\\ learning is transferred from\\ UP-256-512}\\
   \hline
   \thead{Output-layer: ‘O’\\ Size: 1024} & \thead{Decode input from ‘D’\\ up scaled to 1024 units.\\ And is compared with\\ linearized input\\ lattice of 32x32}  &\thead{Decode input from ‘D’\\ up scaled to 1024 units.\\ And is compared with\\ linearized input\\ lattice of 32x32} &\thead{Decode input from ‘D’\\ up-scaled to 1024 units.\\ Not trainable as\\ learning is transferred from\\ UP-512-1024}\\
   \hline
  \end{tabular}
  \caption{\label{table:NNs} A comparison between techniques utilized in training various NNs.}
  }
\end{table}


\end{document}